\documentclass{aa}    
\input{epsf}
\usepackage{graphicx}
\newcommand{\be}{\begin{eqnarray}}    
\newcommand{\ee}{\end{eqnarray}}     
\begin{document}     
      
\title{Statistical characteristics of Large Scale Structure}     
         
\author{M. Demia\'nski \inst{1,2}, 
\and A.G. Doroshkevich \inst{3,4} } 
\offprints{}
\institute{Institute of Theoretical Physics, University of Warsaw,    
00-681 Warsaw, Poland\\     
\and Department of Astronomy, Williams College,     
Williamstown, MA 01267, USA\\     
\and Theoretical Astrophysics Center,     
Juliane Maries Vej 30,     
DK-2100 Copenhagen \O, Denmark\\     
\and Keldysh Institute of Applied Mathematics,    
Russian Academy of Sciences, 125047, Moscow,  Russia    
}     

\date{Received 2002  .../ Accepted ...,     }     
          
\abstract{ 
We investigate the mass functions of different elements of the Large
Scale Structure -- walls, pancakes, filaments and clouds -- and
the impact of transverse motions -- expansion and/or compression --
on their statistical characteristics.  Using the Zel'dovich theory 
of gravitational instability we show that the mass functions of all
structure elements are approximately the same and the mass of all
elements is found to be concentrated near the corresponding mean
mass. At high redshifts, both the mass function and the mean mass of
formed elements depend upon the small scale part of the initial power
spectrum and, in particular, upon the mass of dominant fraction of
dark matter (DM) particles. These results generalize the
Press-Schechter approach and are used to obtain independent estimates
of probable redshifts of the reionization and reheating periods of 
the Universe.
We show that the transverse motions do not significantly change the
redshift evolution of the observed mass function and the mean linear
number density of low mass pancakes related to absorption lines in the
spectra of the farthest quasars. We apply this approach to the
observed Lyman-$\alpha$ clouds and obtain direct estimates of the
variance of initial density perturbations and the shape of initial
power spectrum on small scale. In turn, these estimates restrict the
mass of dominant fraction of DM particles.
  }   
  
\maketitle     
      
\section{Introduction} 

In recent years numerical simulations are being more often used to
investigate the process of formation and evolution of the Large Scale
Structure (LSS) of the Universe. This trend is strongly stimulated by
the rapid growth of computer facilities and it provides better and
better high quality simulations (see e.g., Benson et al. 2001; Smith
et al. 2002; Frenk 2002).  Modern simulations performed in large boxes
and with high resolution can take into account simultaneous action of
many important physical factors and they substantially increase
available information about the process of LSS formation and
evolution.
       
At small redshifts the LSS is observed in the large galaxy 
surveys as a system of filaments and walls. At high redshifts the LSS
is observed mainly as the Lyman-$\alpha$ clouds identified through
absorption lines in spectra of the farthest quasars. All these 
elements of the LSS are quite well
reproduced in numerical simulations what indicates their close
relation with the initial power spectrum. However, this relation is
not yet clear and a quantitative description of the LSS elements
only recently got under way. Now the analysis of both simulated and
observed catalogues of galaxies is focused on the discussion of the
correlation function, $\xi(r)$, and the power spectrum, $p(k)$, while
other characteristics of the LSS are not determined.  This limited
description demonstrates that numerical simulations cannot substitute
theoretical models of structure formation which provide the basis for
much more detailed quantitative description of both observed and
simulated large scale matter distribution.

Theoretical models reveal the main factors that influence the process 
of structure formation and evolution and clarify links between them and
the measured characteristics of the LSS. This opened up a possibility
of formulating new and promising approaches of statistical description
of both simulated and observed characteristics of the LSS such as, for
example, a set of mass functions of structure elements, their
separations and so on. These characteristics can be measured with such
powerful methods as the Minimal Spanning Tree (MST) technique (Barrow,
Bhavsar \,\&\,Sonoda 1985, van de Weygaert 1991; Demia\'nski et al.
2000 hereafter DDMT; Doroshkevich et al. 2001) and the Minkowski 
Functional method (see, e.g., Kerscher 2000) among others. The well 
known example of this approach is provided by the Press--Schechter 
relation for the mass function of high density clouds which has 
been widely used and discussed during the last thirty years (see, 
e.g., Loeb and Barkana 2001; Sheth and Tormen 2002, Scannapieco 
and Barkana 2002).

Some statistical characteristics of the LSS has been derived 
in our previous papers (Demia\'nski \& Doroshkevich 1999, 
hereafter DD99; DDMT). They
are based on the nonlinear theory of gravitational instability 
(Zel'dovich 1970; Shandarin\,\&\,Zel'dovich 1989) applied to 
a CDM--like broad band initial power spectrum of perturbations. 
This approach allows to outline the general tendencies of the 
LSS evolution and demonstrates a leading role of the initial 
velocity field and of the successive merging of structure 
elements in formation of the observed and simulated large scale 
matter distribution. It links some quantitative characteristics 
of galaxy walls with the initial power spectrum. 

Comparison of the theoretical expectations with measured 
statistical parameters of the most conspicuous wall-like 
component of structure was performed in DDMT and in Doroshkevich 
et al. (2002) for the mock 2dF survey (Cole et al. 1998), the 
SDSS Early Data Release (Stoughton et al. 2002), the Las 
Campanas Redshift Survey (Shectman et al. 1996) and the 
Durham/UKST Redshift Survey (Ratcliffe et al. 1998). This analysis 
confirms that the walls are gravitationally confined and 
partly relaxed Zel'dovich pancakes formed presumably 
due to the 1D collapse of matter. Such interpretation was 
already proposed in Thompson \& Gregory (1978) and Oort 
(1983) just after observations of the first walls. 

From this comparison it follows that the main measured characteristics
of walls are quite well expressed in terms of the time scale, the
coherent length and correlation functions of the initial velocity
field, set by the power spectrum of initial perturbations. The time
scale and the coherent length are expressed through the spectral
moments and amplitude of initial perturbations, and the basic 
parameters of the cosmological model. This analysis provides 
independent estimates of the amplitude of initial perturbations 
with a scatter $\leq$20\% and verifies their Gaussianity.

In this paper we extend the model considered in DD99 and DDMT by
taking into account the action of factors responsible for the
evolution of structure elements after their formation. Among the 
most important are: the expansion and compression of LSS elements 
in the transverse directions, the small scale damping of initial
perturbations caused by the random motions of DM particles and 
Jeans damping, and the acceleration of cloud formation within 
larger structure elements, what creates a large scale bias between 
the spatial distribution of DM and luminous matter. As before, 
this analysis is based on the Zel'dovich theory of gravitational 
instability which describes the process of LSS formation as a 
successive compression of matter along three orthogonal directions. 
This implies a successive transformation of pancakes into filaments 
and filaments into clouds with a progressive growth of the masses 
and sizes of structure elements due to their merging.

First of all, this approach allows one to estimate analytically the
rates of formation of LSS elements and their mass functions in a wide
range of redshifts. For all LSS elements -- pancakes, filaments and
high density clouds -- the mass functions are found to be quite
similar to each other. They are also similar to those predicted by 
the Press--Schechter formalism and its extension for elliptical 
clouds (Sheth and Thormen 2002). This similarity demonstrates the 
generic dependence of the characteristics of the LSS on the initial 
power spectrum and indicates that the shape of collapsed clouds 
influences the rate of collapse but does not change significantly 
the mass functions. These results complement the Press--Schechter 
approach and allow one to obtain independent estimates of the mass 
functions for a wider class of objects.

A special problem is the interaction of small and large scale
perturbations. As is well known, it accelerates or decelerates
formation of clouds within compressed or expanded LSS elements. 
In simulations this interaction manifests itself by a strong
concentration of halos within filaments and walls at all redshifts. 
In observations it is seen as a strong concentration of galaxies 
within the richer filaments and walls. In contrast, a significant 
fraction of both DM and baryons remains within low mass pancakes 
which are more homogeneously distributed than the luminous matter.  
The approach used in this paper allows one to quantify this 
interaction.  We show that the formation of high density halos 
and galaxies is modulated by the large scale initial velocity 
field what explains qualitatively the large scale bias.

Using our approach we can describe the basic properties of 
long--lived pancakes observed as the Ly-$\alpha$ forest in a wide 
range of redshifts. In this paper we discuss the evolution of two 
most important characteristics of such pancakes: their mass 
function and their mean number density along the line of sight. 
Both characteristics are evidently changing with redshift due to 
the formation of new and merging of old pancakes, and their 
transverse compression and/or expansion. These characteristics 
depend also upon the cutoff of the initial power spectrum at 
small scale.

Both characteristics play a key role in the  discussion of observed
evolution of the Ly-$\alpha$ forest and we used them (Demia\'nski, 
Doroshkevich\,\&\,Turchaninov 2003) to show that the observed 
properties of absorbers are consistent with theoretical expectations.  
These results restrict also the spectral moments and the mass of dark 
matter (DM) particles to $M_{DM}\geq$ 1--5 keV.  This approach 
can be also used to measure the shape of initial power spectrum down 
to scales of $\sim 30h^{-1}$kpc (Demia\'nski\,\&\,Doroshkevich 2003).  
       
This paper is organized as follows: In Secs. 2 basic relations        
are introduced. In Sec. 3 the redshift dependence of the expected       
matter fraction assigned to various types of structure elements       
is found. In Sec. 4 we discuss the interaction of large and small 
scale perturbations. In Sec. 5 the joint mass functions of DM 
structure elements are considered. In Secs. 6\,\&\,7 we discuss 
the redshift evolution of statistical characteristics of filaments 
and pancakes. Short conclusion can be found in Sec. 8. Some technical 
details are given in Appendixes A \& B.       
       
\section{Basic statistical characteristics of the Zel'dovich 
theory}       
       
In this section we present the basic statistical characteristics        
of Zel'dovich approximate nonlinear theory of gravitational
instability used later to         
describe the process of structure formation and evolution. 
Main ideas and characteristics were already introduced in DD99 
and are repeated here without discussion. Some definitions are 
improved and corrected what makes the approach more transparent.          
       
In the Zel'dovich theory (Zel'dovich 1970; Shandarin \& Zel'dovich 
1989) the Eulerian, $r_i$, and the Lagrangian, $\tilde{q}_i$, 
coordinates of particles (fluid elements) are related by 
\be       
r_i = (1+z)^{-1}[\tilde{q}_i - B(z)S_i(\tilde{{\bf q}})]\,,       
\label{eq1}       
\ee       
where $z$ denotes the redshift, $B(z)$ describes growth of       
perturbations in the linear theory, and the gradient vector       
$S_i(\tilde{\bf q})=\partial \phi/\partial \tilde{q}_i$ 
characterizes the spatial distribution of perturbations. The 
Lagrangian coordinates of a particle, $\tilde{q}_i$, are its 
unperturbed coordinates in the real space, $r_i(z=0) = 
\tilde{q}_i$. For the spatially flat $\Lambda$CDM cosmological 
model the function $B(z)$ can be approximated (DD99) as follows:       
\be       
B^{-3}_\Lambda(z)\approx {\Omega_\Lambda+2.2\Omega_m(1+z)^3\over       
1+1.2\Omega_m},~~\Omega_\Lambda+\Omega_m=1\,.		       
\label{B1}       
\ee       
Here $\Omega_m \& \Omega_\Lambda$ are the dimensionless matter 
density and cosmological constant. This fit is reasonably accurate 
for all $z$ and is normalized at $z=0$ by the condition $B(0)=1$.       
       
In this paper we consider only power spectra with the Harrison        
-- Zel'dovich asymptotic, $p(k)\propto k$, at $k\rightarrow 0$, 
and CDM-like or WDM-like transfer functions, $T^2(k)$, introduced 
in Bardeen et al. (1986, hereafter BBKS):        
\be       
p(k)={ A^{2}k\over 4\pi k_0^4}T^2\left({k\over k_0}\right),\quad 
k_0=\Omega_{m} h^{2}{\rm Mpc}^{-1},
\label{spc}       
\ee       
where $A$ and $h=H_0/100$ km/s/Mpc are the dimensionless amplitude 
of perturbations and the Hubble constant, $k$ is the comoving wave 
number. The same approach can be applied for arbitrary initial power 
spectra.        
       
\subsection{Coherent lengths and correlation functions       
of initial density and velocity fields}       
       
For the spectra (\ref{spc}), the coherent lengths of initial density        
and velocity (or displacement) fields, $l_\rho$ and $l_v$, are        
expressed through the spectral moments, namely, $m_{-2}$ and $m_0$,        
(DD99):       
\[      
m_0=\int_0^\infty x^3T^2(x)D_W(x)D_J(x)dx,
\]
\[
m_{-2}=\int_0^\infty xT^2(x)dx\approx 0.023,\quad x=k/k_0\,,
\]       
\be       
l_v ={1\over k_0\sqrt{m_{-2}}}\approx {6.6\over {\Omega_{m}h^{2}}}        
{\rm Mpc},\quad l_\rho=q_0l_v={5\over k_0}{m_{-2}^{3/2}\over m_0}\,.       
\label{lv}       
\ee       
Here $D_W$ and $D_J$ describe damping of perturbations caused by the random 
motions of DM particles (see BBKS) and suppression of 
formation of baryonic clouds due to gaseous pressure (Jeans 
damping, see, e.g., Matarrese\,\&\,Mohayaee 2002), respectively. 
Analysis of simulations and observed Ly-$\alpha$ forest (Narayanan 
et al. 2000; Barkana, Haiman \& Ostriker 2001; Demia\'nski, 
Doroshkevich \& Turchaninov 2003) allows to estimate $q_0$ and 
$M_{DM}$ as follows:  
\be 
M_{DM}\geq 1-5{\rm keV},\quad q_0\simeq (0.5 - 1)\cdot 10^{-2}\,.
\label{dm}
\ee      
The characteristic masses of DM clouds associated with        
the coherent lengths $l_{v}$ and $l_{\rho}$ are       
\be       
M_v = {4\pi\over 3}<\rho> l_v^3\approx {2\cdot 10^{14}       
\over (\Omega_mh^2)^2}M_\odot,\quad M_\rho = q_0^3M_v\,.       
\label{mv}       
\ee       
       
Large difference between the coherent lengths $l_v~\&~l_\rho$        
indicates that formation of galaxy filaments and walls         
observed at small redshifts is mainly driven by the larger 
coherent length, $l_v$. The small scale correlations of 
density field characterized by the coherent length $l_\rho$ 
are more important during formation, at high redshifts, of low 
mass objects such as the first population of galaxies and 
Ly-$\alpha$ absorbers.       
       
For spectra under consideration, the normalized correlation       
functions of displacement and density fields,        
\be       
\xi_v(q)=3{\langle({\bf q\cdot S}(\tilde{\bf q}_1))({\bf q\cdot S}       
(\tilde{\bf q}_2))\rangle\over \sigma_s^2q^2},\quad
\sigma_s^2={A^{2}m_{-2}\over k_0^2}\,,  
\label{vrho}    
\ee       
\[       
\xi_\rho(q)={\langle\rho(\tilde{\bf q}_1)\rho(\tilde{\bf q}_2)       
\rangle-\langle\rho\rangle^2\over\langle\rho\rangle^2       
\sigma_\rho^2},	\quad \sigma_\rho^2=A^{2}m_0
\]   
\[
{\bf q}=(\tilde{\bf q}_1-\tilde{\bf q}_2)/l_v,\quad 
q=|\tilde{\bf q}_1-\tilde{\bf q}_2|/l_v\,,
\]    
were approximated in DD99. Here, as before, $\tilde{\bf q}_1 
~\&~ \tilde{\bf q}_2$ are real unperturbed coordinates of two 
particles at $z=$ 0, $\sigma_s^2$ and $\sigma_\rho^2$ 
are variances of the displacement and the density fields, and 
$A$ is the amplitude of initial perturbations introduced in 
(\ref{spc}). Due to homogeneity and isotropy of both the 
background matter distribution and perturbations the correlation 
functions depend only on the differences of coordinates $|{\bf q}|$. 
For the most interesting case $q_0\ll$ 1, $q\leq$ 1, the 
correlation functions can be written as follows:       
\be       
\xi_v\approx 1-{q^2\over\sqrt{q^2+q^2_0}}{q_0+2\sqrt{q^2+q^2_0}       
\over q_0+\sqrt{q^2+q^2_0}}\,,       
\label{xiv}       
\ee       
\be       
\xi_\rho=-{q_0\over 15}{1\over q^4}{d\over dq}q^4{d\xi_v\over dq}       
\approx {q_0\over\sqrt{q_0^2+q^2}},~       
{\rm for} ~q\leq q_0\ll 1\,,       
\label{xirho}       
\ee       
and $\xi_\rho\ll$ 1, for $q_0\leq q$.        
At $q\ll q_0$ both functions, $\xi_v\,\&\,\xi_\rho$, only weakly 
depend upon the higher spectral moments, $m_2, m_4, ...$. Some 
of the more cumbersome correlation and structure functions of 
perturbations are discussed in DD99 and are presented in the 
Appendix A.        
       
\subsection{Basic relations}       
       
As was shown in DD99, to describe the formation of structure        
elements the basic equation (\ref{eq1}) has to be rewritten using        
the differences of particle coordinates and displacements. 
The separation of two particles with Lagrangian and Euler 
coordinates $\tilde{\bf q}_1\,\&\,\tilde{\bf q}_2$, and        
${\bf r}_1\,\&\,{\bf r}_2$, respectively, is described by the 
equations (i=1,2,3):       
\be       
\Delta r_i=({\bf r}_1-{\bf r}_2)_i ={l_v\over 1+z}\left[q_i-
{B\Delta S_i(q_i)\over l_v}\right] \,,
\label{eq2}       
\ee   
\[
\Delta S_i=S_i(\tilde{\bf q}_1/l_v)-S_i(\tilde{\bf q}_2/l_v),\quad 
q_i={(\tilde{\bf q}_1-\tilde{\bf q}_2)_i\over l_v}\,.
\]

Evidently, for Gaussian initial perturbations, the PDF of 
$\Delta S_i$ is also Gaussian with
\be 
\langle\Delta S_i(q_i)\rangle =0,\quad {3\over\sigma_s^2}
\langle\Delta S_i^2(q_i)\rangle =\sigma_q^2=
2[1-\xi_v(q_i)]\,.
\label{mns}
\ee   
These relations show the symmetry between expansion and 
compression of matter and the progressive growth of 
$\langle\Delta S_i^2(q_i)\rangle\propto |q_i|$ for larger 
$|q_{i}|$ ($1\geq |q_i|\gg q_0$). These expressions indicate 
also a relatively slower rate of evolution of objects 
with larger $|q_i|$, because it is defined by the ratio 
$\Delta S_i/|q_i|\propto\sigma_q/q\propto q^{-1/2}$.   

According to the relations (\ref{eq2}), when two particles 
with different Lagrangian coordinates $\tilde{\bf q}_1$ and 
$\tilde{\bf q}_2$ meet at the same Eulerian point $\bf r$ a 
caustic -- Zel'dovich pancake -- with the surface mass density 
$<\rho>|\tilde{\bf q}_1-\tilde{\bf q}_2|$ forms. Following 
DD99 we assume that {\it all} particles situated between 
these two boundary particles are also incorporated into 
the same pancake. This assumption is also used in the 
adhesion approach (see, e.g., Shandarin \& Zel'dovich 1989).        
Comparison of statistical characteristics of pancakes with        
simulations (DD99; DDMT) verifies this approach and shows        
that long lived and partly relaxed richer walls accumulate a 
significant fraction of compressed matter.
The same condition can be used 
to describe formation of filaments and clouds in the course 
of successive collapse along two and three axes. 

For the general description of structure formation it is 
convenient to combine the three equations (\ref{eq2}) into 
one scalar equation, namely,
\be
q_i\Delta r_i ={l_vq^2\over 1+z}[1-\tau(z)F({\bf q})]\,,
\label{eq3}
\ee
\be       
\tau(z)={\sigma_s\over\sqrt{3}l_v}B(z) = \tau_0B(z),\quad
F({\bf q})={\sqrt{3}{\bf q}\Delta {\bf S}({\bf q})\over 
\sigma_sq^2}\,.    
\label{tau}       
\ee
The random function $F({\bf q})$ with the Gaussian PDF and 
\be
\langle F\rangle=0,\quad\langle F^2\rangle=\sigma_F^2({\bf q}), 
\label{eq4}
\ee
characterizes the evolution of a region defined by the given limits of
${\bf q}$. The dispersion $\sigma_F({\bf q})$ is expressed through the
structure functions introduced in Appendix A and DD99. The conditions
$F({\bf q})\geq 1$ and $F({\bf q})\leq 0$ separate the collapsed and
expanded parts of the volume, the condition $1\geq F({\bf q})\geq 0$
separates part of the volume that will collapse later. This approach 
allows to obtain main characteristics of structure. However, its 
practical applications are not simple due to complicated form of 
the function $\sigma_F( {\bf q})$.

\subsection{Characteristics of the deformation field}       
       
In DD99 we assumed that the deformation field is dominated by        
two lowest harmonics what allows to characterize the formation        
of structure by three weakly correlated components of the      
displacements. Analysis performed in Appendix B shows that,        
for the spectra (\ref{spc}), this assumption is valid with        
a precision better than 10\%. For the most interesting cases        
$q_0\ll q\ll$1 and $q\ll q_0\ll$1, 
the normalized         
amplitudes of several lowest spherical harmonics of the deformation 
field, $b_l^2$, are given by:       
\[       
b_0^2\approx 0.533,~~ b_2^2\approx 0.381,~~ b_4^2\approx       
0.037,~~ b_6^2\approx 0.014, ...	       
\]       
\[b_0^2\approx 0.55,\quad b_2^2\approx 0.44\gg b_4^2,~~       
b_6^2 ... .			       
\]       
and the contribution of higher order harmonics with $l\geq$ 
4 is only $\sim$ 1\% for low mass clouds with $q\leq q_0\ll$ 
1, and it reaches $\sim$ 8\% for more massive clouds with $q_0\ll 
q\leq$ 1.        
       
These results justify the assumption made in DD99 to neglect       
higher order harmonics of perturbations with $l\geq 4$ and they
confirm that the formation of structure can be approximately        
described by the spherical and quadrupole components of the 
deformation field. The influence of higher harmonics, even with 
small amplitude, leads to a small scale disruption of the compressed 
clouds because of the strong instability of thin pancake-like 
condensations and, so, to the formation of internal structure 
of clouds. 

These results indicate also that when the process of 
structure formation is described by the function $F({\bf q})$ the 
ellipsoidal or, in the case of two dimensional problem, elliptical 
volumes/areas are preferable. 

\subsection{Simple approximation}
       
When only two spherical harmonics are taken into account, 
the general deformation of any cloud can be described by 
its deformations along the three orthogonal principal axes, 
namely, $x_1, x_2, \& x_3$. In this case, we can use a 
simpler approach and consider again the three equations (\ref{eq2}) 
instead of (\ref{eq3}). Applying this approach it is possible to obtain some 
approximate characteristics of the LSS. However, its abilities 
are restricted and some important problems can be solved only 
with the general approach (\ref{eq3}). 

The numbering of principal axes is arbitrary but, further 
on, we will usually assume that        
\be       
\Delta S_1/q_1\geq\Delta S_2/q_2\geq\Delta S_3/q_3\,,
\label{s123}
\ee        
and the cloud collapses fastest along the first        
axis whereas slower collapse -- or even expansion takes        
place along the third axis. For $q_i\rightarrow 0$ this choice 
agrees with the ordering of principal axes of the deformation 
tensor (Zel'dovich 1970; Shandarin \& Zel'dovich 1989).        
       
The correlations of differences of displacements along the 
principal axes are relatively small (DD99),       
\be       
r_{ij}= {3\langle\Delta S_i(q_i)\Delta S_j(q_j)\rangle\over       
\sigma_s^2\sigma_q(q_i)\sigma_q(q_j)}\leq {1\over 3},~~ i\neq j\,,       
\label{rij}       
\ee       
what allows us to consider, in many problems, the components 
of the deformation field as uncorrelated. Dispersions $\sigma_s
\,\&\,\sigma_q$ were defined in (\ref{vrho}\,\&\,\ref{mns}).

For quantitative estimates it is convenient to rewrite 
(\ref{eq2}) in a dimensionless form       
\be
\Delta r_i= {l_vq_i\over 1+z}\left[1-{\tau(z) s_i\over
\mu(q_i)}\right],\quad s_i={\sqrt{3}\Delta S_i\over\sigma_q
\sigma_s}\,,      
\ee 
\be       
\mu(q)={q_i\over\sigma_q}={q_i\over\sqrt{2[1-\xi_v(q_i)]}},
\quad \eta(q_i,z)={\mu(q_i)\over\sqrt{2}\tau(z)}\,.      
\label{eta1}
\ee
The probability distribution function (PDF) of dimensionless
differences of displacements $s_{i}$ is Gaussian and it is given by:
\be       
{d^3W(s_1,s_2,s_3)\over ds_1 ds_2 ds_3}
=6(2\pi)^{-3/2} \exp[-0.5(s_1^2+s^2_2+s_3^2)]\,.       
\label{ws}       
\ee    
Here the factor 6 appears because of the imposed restrictions
(\ref{s123}) on the range of variables.

As is seen from (\ref{xiv},\,\ref{eta1}), for two limiting 
cases $q_i\gg q_0$ and $q_i\ll q_0$, we have     
\be       
\mu(q)\approx {\sqrt{q}\over 2},~~\eta={1\over 2\tau}       
\sqrt{q\over 2},~~ q_0\ll q\ll 1,       
\label{mu1}       
\ee       
\be       
\mu(q)=\mu_0\approx \sqrt{q_0\over 3},~\eta=\eta_0=       
{1\over \tau}\sqrt{q_0\over 6},~~ q\ll q_0\ll 1,       
\label{mu2}       
\ee       
and the two particles under consideration moving along the $i^{th}$
axis will cross when $s_i\geq \sqrt{2}\eta_i$.  For smaller $q_i\leq
q_0\ll$ 1 the condition of the crossing does not depend on the size of
the pancake, it means that collapse of all small pancakes takes the 
same amount of time.

The parameter $q_0$ (\ref{lv}, \ref{xirho}), characterizes the 
damping scale in the initial power spectrum and discriminates 
regions of strong and small correlations of the initial density 
and velocity fields. As is seen from (\ref{mu2}), the condition 
$\eta_0=1$ introduces also the typical redshift, $z=z_r$,
\be
\quad \tau(z_r)=\tau_r=\sqrt{q_0\over 6},\quad z_r\approx 1+ 
{\tau_0\over\tau_r}\left({1+1.2\Omega_m\over 2.2\Omega_m}
\right)^{1/3}\,.
\label{zr}
\ee
At larger redshifts, $z\geq z_r$, $\tau\leq \tau_r$, the 
formation and evolution of the LSS elements strongly 
depends on the small scale correlations, while later on 
their influence progressively decrease. For $\Omega_m=0.3, 
q_0=10^{-2}$ and $\tau_0\sim$ 0.3 we have $z_r\approx 6.5$ 
what coinsides with the observed low limit of the redshift 
of reionization of the Universe (Fan et al. 2001).

\section{Fraction of matter accumulated by structure elements}       

The fractions of matter accumulated by pancakes and filaments 
were found in DD99. Here we extend this approach and introduce 
three cumulative distribution functions for more 
detailed description of fractions of matter accumulated by 
pancakes, filaments and walls with different sizes and rate 
of expansion and/or compression. In turn, this extension allows one 
to obtain the mass functions of all three kinds of LSS 
elements and to take into account the further evolution of pancakes 
after their formation. These problems are discussed in Secs. 5 \& 7.  
       
\subsection{Cumulative distribution functions of 
displacement differences}       
       
As is seen from (\ref{eta1}), the conditions $\Delta r_1\leq 0$ 
and $s_1/\sqrt{2}\geq \eta_(q_1)=\eta_1$ are equivalent 
and both define collapse of a cloud along the axis of 
the most rapid compression. To discriminate between pancakes, 
filaments and walls, we have to use also conditions which 
restrict the motion along two other directions. For more 
detailed description of pancakes we have also to restrict 
the rate of pancake expansion. These conditions are consistent with 
restrictions (\ref{s123}) and they can be summarized as follows:
\be       
{s_1\over\sqrt{2}}\geq\zeta_1,\quad \zeta_1\geq\zeta_2
\geq{s_2\over\sqrt{2}}\geq{s_3\over\sqrt{2}},\quad        
\zeta_3\geq{s_3\over\sqrt{2}}\geq -\infty\,,
\label{cp}     
\ee
where the parameters $\zeta_{i}$ restrict the rate of deformation in
different directions.

The first condition in (\ref{cp}) assumes that for a given 
$\tau(z)$ the clouds with $\zeta_1=\eta_1\geq\eta_0$, 
$\mu(q_1)\geq\sqrt{2}\zeta_1\tau$ are collapsed along the 
first axis. For $\zeta_2=\eta(q_2)\geq\eta_0,\,\zeta_3=
\eta(q_3)\geq\eta_0$, the next two conditions exclude from the 
consideration collapsed  filaments with $\mu(q_2)\geq\sqrt{2}
\zeta_2\tau$ and clouds with $\mu(q_3)\geq\sqrt{2}\zeta_3\tau$. 
For $\zeta_3\leq\zeta_2\leq 0$ they restrict the expansion rate of
expanding pancakes.

Similar conditions, namely,
\be       
{s_1\over\sqrt{2}}\geq\zeta_1,\quad {s_1\over\sqrt{2}}\geq 
{s_2\over\sqrt{2}}\geq \zeta_2\geq\zeta_3\geq{s_3\over\sqrt{2}}
\geq -\infty
\label{cf}     
\ee       
\be       
{s_1\over\sqrt{2}}\geq\zeta_1,\quad {s_1\over\sqrt{2}}\geq 
{s_2\over\sqrt{2}}\geq\zeta_2,
\quad {s_2\over\sqrt{2}}\geq {s_3\over\sqrt{2}}\geq\zeta_3\,,       
\label{ccl}     
\ee       
with $\zeta_1=\eta_1\geq\zeta_2=\eta_2\geq\eta_0$, restrict 
the minimal sizes of filaments and, for $\zeta_1=\eta_1\geq
\zeta_2=\eta_2\geq\zeta_3=\eta_3\geq\eta_0=\eta(q=0)$, of 
clouds collapsed at a given $\tau(z)$ . 

Integration of the PDF (\ref{ws}) under conditions (\ref{cp},
~\ref{cf}\,\&\,\ref{ccl}) results in the cumulative PDFs  
characterizing the evolution of the LSS elements:
\be       
W_p(\zeta_1,\zeta_2,\zeta_3) = {3\over 8}(1-e_1)(1+e_3))
(1+2e_2-e_3)\,,       
\label{wp}       
\ee       
\be       
W_f(\zeta_1,\zeta_2,\zeta_3) = {3\over 8}(1-e_1)(1+e_3)
(1+e_1-2e_2)\,,       
\label{wf}       
\ee       
\be       
W_{cl}(\zeta_1,\zeta_2,\zeta_3) = {1\over 8}(1-e_1)[1+e_1+
e_1^2-3e_3+
\label{wcl}       
\ee       
\[
3e_3(e_3-e_1)-3(e_2-e_3)^2]\,,       
\]
where $e_i$=erf$(\zeta_i)$. 

Functions $W_p, W_f\,\&\,W_{cl}$ describe the fractions of matter 
assigned to pancakes, filaments and clouds, respectively, 
under conditions (\ref{cp},~\ref{cf}\,\&\,\ref{ccl}). 
For $\zeta_2=\zeta_3=\eta_2,\,e_2=e_3$, expressions (\ref{wp}, 
\ref{wf}) become identical to those obtained in DD99.  

\subsection{Maximal fraction of matter accumulated by 
structure elements}       
       
From the relations (\ref{wp}, \ref{wf}, \ref{wcl}) it follows 
that in the Zel'dovich theory the maximal fractions of matter 
which can be accumulated by clouds, filaments and pancakes, 
for $q_i=0$, $\eta_i=\eta_0=\tau^{-1}(z)\sqrt{q_0/6}$, 
are
\be       
W_{cl}\approx {1\over 8}{\rm erfc}^3(\eta_0),\quad 
W_f={3\over 8}{\rm erfc}^2(\eta_0)(1+{\rm erf}(\eta_0))\,,      
\label{www}       
\ee  
\[
W_p\approx {3\over 8}{\rm erfc}(\eta_0)(1+{\rm erf}(
\eta_0))^2\,. 
\]     
We see that during the early evolutionary stages, when 
$z\gg z_r,~ \tau\ll\tau_r, ~\eta_0\gg 1$, the
fractions (\ref{www}) increase $\propto \exp(-\kappa
\eta_0^2)$ with $\kappa=1, 2, 3$ for pancakes, filaments 
and clouds. During later evolutionary stages, when 
$z\ll z_r,~ \tau\gg\tau_r,~ \eta_0\ll 1$,
we get from (\ref{www}):
\[
W_{cl}\approx 0.125,\quad W_f\approx W_p\approx 0.375\,.
\]

In the Zel'dovich theory motion of matter along the three principal
directions as described by the equation (\ref{eq2}) is considered to
be independent. It means that the interaction of matter is neglected
and therefore in this approach the matter fraction accumulated by
filaments and clouds and the rate of collapse are usually underestimated.
For small $\tau$, the acceleration of the collapse along the $x_2$ 
and $x_3$ axes after the collapse along the $x_1$ axis and the 
interaction of small and large scale perturbations increase the fraction 
of matter accumulated by the collapsing cloud by a factor of about 
2. The same effects also shift the boundary between  
structure elements and increase the final fraction of matter 
assigned to filaments and clouds. The popular Press-Schechter 
approach faces the same problems because the assumption of 
spherical collapse overestimates the collapse rate and the 
fraction of collapsed matter. 

\begin{figure}       
\centering       
\epsfxsize=7.5 cm       
\epsfbox{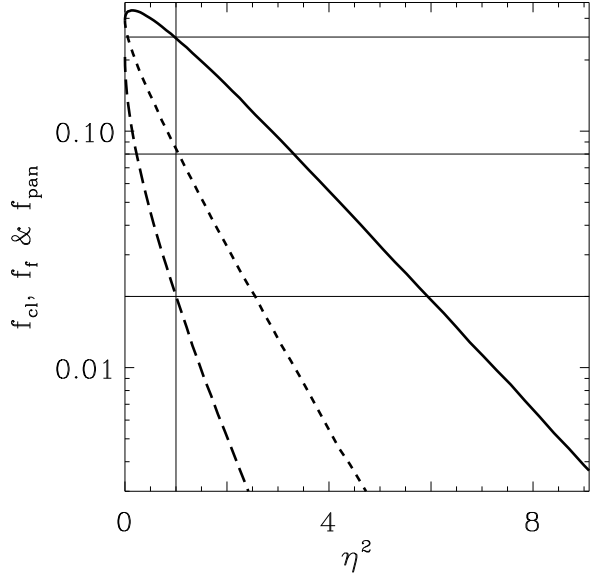}       
\vspace{1.2cm}       
\caption{Fraction of the DM component accumulated by pancakes 
and walls (solid line), filaments (dashed line) and clouds (long 
dashed line) plotted versus $\eta^2$. These fractions at the 
moment $\eta =$ 1, $\tau=\tau_r$, are shown by thin straight lines.         
}   
\label{fig1}    
\end{figure}

The cross correlations of orthogonal displacements as given 
by (\ref{rij}), though small, increase the maximal fraction of 
matter accumulated by clouds and filaments up to $W_{cl}\sim 
0.2$ and $W_{f}\sim 0.3$, respectively, and decrease the 
fraction of matter accumulated by pancakes to $W_{p}\sim 0.3$. 
These maximal fractions are plotted in Fig. \ref{fig1} versus 
$\eta^2/\eta_0^2$ for all three types of structure elements. 
As it is apparent from Fig. \ref{fig1}, already at $\tau=\tau_r$, 
$\eta=\eta_0$, $\sim$ 25\%, 8\% and 2\% of matter is accumulated 
by pancakes, filaments and clouds, respectively. At $\tau
\leq \tau_r$, these mass fractions increase $\propto 
\exp(-\kappa\eta^2)$, $\kappa=1, 2, 3$ what coincides with 
(\ref{www}). 

These estimates of the mass fractions are quite reasonable for 
large redshifts, when the formation of structure elements from 
the dispersed matter dominates. At small redshifts the merging 
of the LSS elements distorts these estimates and, for example, 
some fraction of filaments and clouds is incorporated into richer 
walls. These estimates can be improved by a comparison with high 
resolution simulations and observations of Ly-$\alpha$ forest 
at high redshifts. 

\subsection{Distribution function of the sizes of structure 
elements}
        
The relations (\ref{wp}, \ref{wf}, \ref{wcl}) are monotonic 
functions of $\zeta_i$ and, for $\zeta_i=\eta_i\geq\eta_0$, 
they represent the cumulative three parameters distribution 
functions of structure elements with respect of their sizes 
measured by differences of Lagrangian coordinates, $q_1, 
q_2, q_3$, for particles bounding these LSS elements along 
the principal directions. 
This joint PDF for all structure elements is 
\be       
N(q_1,q_2,q_3)\propto \exp(-\eta_1^2-\eta_2^2-
\eta_3^2){d\eta_1\over dq_1}{d\eta_2\over dq_2}
{d\eta_3\over dq_3}        
\label{d3w}       
\ee       
\[       
\eta_0\leq \eta_i\leq\infty,\quad 0\leq q_i\leq\infty\, .       
\] 
Similar relation allowing for the correlations (\ref{rij}) 
can be obtained in the same manner.

The expression (\ref{d3w}) does not discriminate filaments and 
pancakes with respect to their expansion rates described 
in (\ref{wp}) and (\ref{wf}) by negative $\zeta_2$ and 
$\zeta_3$. To allow such discrimination it is necessary to  link 
$\zeta_2\leq 0$ and $\zeta_3\leq 0$ with the size of 
structure elements. The symmetry of the basic characteristics of
structure elements with respect to 
positive and negative $\Delta S_i$ discussed in Sec. 2.2 
and the relations (\ref{mns}) indicate that such an extension can 
be achieved by imposing the following limits in
(\ref{d3w}): 
\[
0\leq q_1\leq\infty,\quad -\infty\leq q_2\leq\infty,
\quad -\infty\leq q_3\leq\infty\,,
\] 
\be
0\leq q_1\leq\infty,\qquad 0\leq q_2\leq\infty,
\quad -\infty\leq q_3\leq\infty\,,
\label{limits}
\ee
\[
0\leq q_1\leq\infty,\qquad 0\leq q_2\leq\infty,
\qquad 0\leq q_3\leq\infty\,,
\] 
for pancakes, filaments and clouds, respectively. Here 
negative $q_2, q_3$ are associated with expanding filaments 
and pancakes. 

\section{Interaction of small and large scale perturbations.        
Large scale bias}       
       
High resolution simulations show that at high redshifts 
substantial fraction of filaments accumulate high density 
halos. In observed redshift surveys galaxies are mainly 
situated within filaments and walls and the population of 
isolated galaxies is quite small. This strong galaxy 
concentration within the LSS elements  -- large scale bias 
-- can be naturally explained by the interaction of 
small and large scale perturbations responsible for the 
formation of galaxies, filaments and walls, respectively. 

As was noted in DD99, the statistical approach allows to  
describe this interaction.  Here we are returning to this 
problem and propose a  simple quantitative measure that
illustrates effects of this interaction. 
We compare the process of formation of two identical 
pancakes with a surface density $m_1=\langle \rho\rangle q_1$, 
and $\mu_1=\mu(q_1)$. We assume that the first -- a reference -- 
pancake is formed at the 'time' moment $\tau_1(z_1)$ and is 
characterized by the function
\[       
\eta(q_1,z_1)= {\mu(q_1)\over {\sqrt{2}\tau_1}}=(1+z_1)
{\mu(q_1)\over \sqrt{2}\tau_0}\,,       
\]       
as given by (\ref{eta1}, \ref{mu2}). For comparison, let us 
consider the process of formation of the second pancake with 
the same size, $q_2=q_1,~\mu(q_2)=\mu_2=\mu_1$, at redshift 
$z_2$ and the successive formation, in the same region, of a 
pancake with $m_3\gg m_2$, $q_3\gg q_2, ~\mu(q_3)=\mu_3\gg
\mu_2$ at redshift $z=z_3\leq z_2$, and at the 'time' moment 
$\tau_3(z_3)\geq\tau_2(z_2)$. We assume that the larger 
pancake with the size $q_3$ accumulates the earlier formed 
smaller pancake with the size $q_2\ll q_3$.        
           
 In this case, the interaction of pancakes can be described by 
replacing $\eta_2$ in (\ref{wp}, \ref{d3w}) with the effective 
parameter $\eta_{eff}$ defined by the relation (DD99):       
\be       
\eta_{eff} = (\eta_2-r_{23}\eta_3)/\sqrt{1-r_{23}^2},	       
\label{eff}       
\ee       
\[       
r_{23}\approx{2\mu_2\mu_3\over q_2q_3}\left[\xi_v\left({q_2-q_3       
\over 2}\right)-\xi_v\left({q_2+q_3\over 2}\right)\right]\,.        
\]       
       
If both pancakes are small, $1\gg q_0\geq q_3\gg q_2$, then 
$r_{23}\rightarrow$ 1, and the strong small scale correlation of 
density field decreases the survival probability of smaller 
pancakes and both pancakes are actually formed at similar 
redshifts $z_2\approx z_3$. However, later formation of a 
larger pancake with $1\gg q_3\geq q_0, ~q_3\gg q_2,~~\mu_3
\gg\mu_2$, accelerates compression of the smaller pancake, 
it increases the effective amplitude of perturbations and shifts 
formation of pancakes  to larger redshifts. For $r_{23}\approx 
\mu_2/\mu_3\ll 1$, and at larger redshifts, when $\eta_i
\propto 1+z$, the general expression (\ref{eff}) can be 
rewritten more transparently as follows:        
\be       
\eta_{eff}\approx \eta_2\left(1-{1+z_3\over 1+z_2}\right) =        
{\mu(q_2)\over\sqrt{2}\tau_0}(z_2-z_3)\,.       
\label{eff2}       
\ee       
Statistical characteristics of pancakes formed at $z=z_1$ and $z=       
z_2$ will be the same for the same value of       
\[       
\eta_{eff}=\eta_1 = (1+z_1){\mu(q_1)\over\sqrt{2}\tau_0}\,.        
\]       
However, as is seen from (\ref{eff2}), this occurs
already at the redshift       
\[       
z_2 = z_1+1+z_3\geq z_1+1\,.       
\]   
 
\begin{figure}
\centering
\epsfxsize=7.cm
\epsfbox{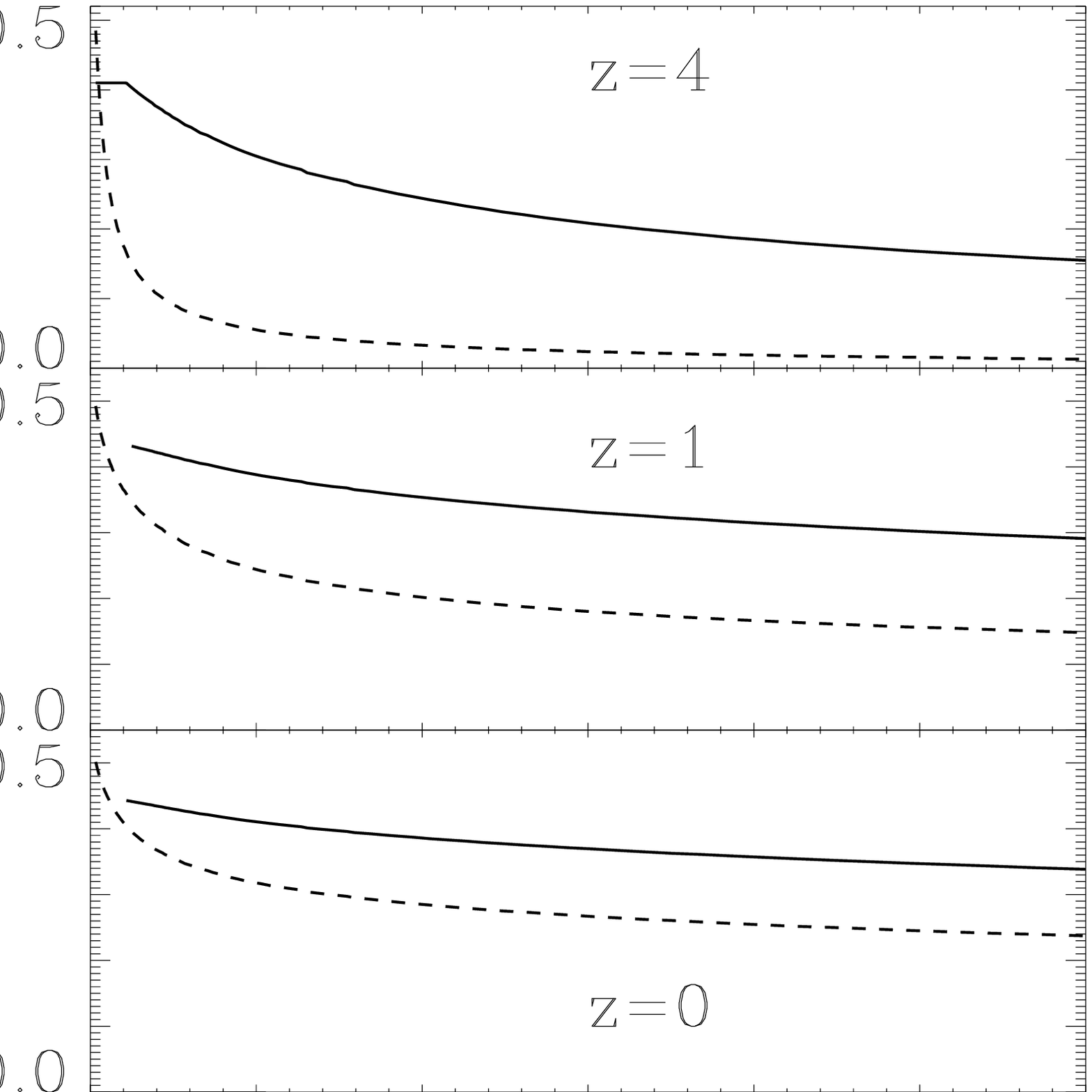}
\vspace{1.2cm}
\caption{Fractions of matter, $f_m$, accumulated by structure elements 
situated within HDRs (solid lines) and LDRs (dashed lines) are plotted 
vs. the threshold overdensity for simulated DM distribution at three 
redshifts.
}
\label{fig2}
\end{figure}
  
This relation gives the quantitative estimate of the influence of        
interaction between the large and small scale perturbations, which          
increases the actual redshift of pancake formation. Pancakes with        
the same size $q_1$ are now formed earlier at redshift $z_2\geq 
z_1$, instead of at $z_1$.        
    
The suppression of pancake formation within expanding regions 
can be considered in the same manner. Thus, for example, for        
{\it negative} $\Delta S_i$ and $-s_i\geq \sqrt{2}\eta(q_i,z)$ 
we have instead of (\ref{eff2})        
\be       
\eta_{eff}\approx \eta_2\left(1+{1+z_3\over 1+z_2}\right)=        
{\mu(q_2)\over\sqrt{2}\tau_0}(2+z_2+z_3)\,,       
\label{eff3}       
\ee       
and from the condition $\eta_{eff}=\eta_1$ we get        
\[       
z_2 = z_1-1-z_3\leq z_1-1\,,       
\]       
what illustrates the influence of this interaction on the moment of
formation of pancakes.        
       
The influence of large scale perturbations on the process of        
formation of small scale objects can be considered as the 
manifestation of large scale bias. Fraction of matter accumulated 
within high density clouds increases rapidly with time (see Fig. 
\ref{fig1}) and halos formed at larger $z$ contain only a small 
fraction of mass of halos formed at smaller $z$. But this bias 
increases the redshift of subclouds formation and their densities, 
promotes the transformation of DM clouds into observed galaxies, 
and makes the internal structure of clouds more complex. Regular 
distribution of young galaxies within high density filaments 
suppresses their feedback and fosters further formation of 
galaxies.     
       
Direct calculation shows that, for larger clouds with the 
correlation coefficient $r_{12}\ll 1$, the influence of the 
nearest collapsed cloud dominates, and the generalization of 
(\ref{eff}\,\&\,\ref{eff2}) to the case of multi step collapse 
at $z_2\geq z_3\geq z_4\geq ...$ practically does not change 
the dominant role of the first term. This means that the multi 
step collapse has to be considered step by step taking into account 
only the next generation of high density condensations. This means        
also that such interactions can be described by the theory 
of Markov processes.   
    
\subsection{Comparison with simulations}     
     
To compare these expectations with simulations we use the high 
resolution simulation (Klypin et al. 1999; Schmalzing et al. 1999) 
performed with adaptive code in a box of (60$h^{-1}$Mpc)$^3$ with 
$N_{tot}=(256)^3$ particles for the Harrison--Zel'dovich primordial 
power spectrum and the BBKS transfer function. This simulation 
approximately corresponds to $q_0\sim 10^{-2}$, $M_{DM}\sim$ 1 
keV. The matter distribution at redshifts $z=4$, $z=1$ and $z=0$ 
were analyzed.     
     
To test the impact of environment on the properties of high      
density clouds the full samples were separated into subsamples of 
high and low density regions (HDRs and LDRs). The subsamples of 
high density regions accumulate $\sim$ 40 -- 45\% of all DM 
particles within richer clouds selected with the friend--of--friend 
method for the threshold density equal to the mean density. Subsamples 
of LDRs were prepared by removing the HDRs from the full samples. 

The fraction of compressed matter is plotted in Fig. \ref{fig2} 
versus the threshold overdensity used for selection of structure 
elements. As is seen from this Figure, at all redshifts the majority 
of highly compressed matter is situated within the HDRs. This excess 
is moderate at $z=0$ and progressively increases with redshift.      
Special test shows that the high density clouds selected at $z=4$      
are equally distributed between HDRs and LDRs selected at $z=0$.      
This fact is consistent with almost equal concentration of 
galaxies within the HDRs and LDRs at $z=0$ in the observed surveys. 
It indicates also that the efficiency of the interaction of small 
and large scale perturbations decreases with redshift. 
          
\section{Mass function and rotation of structure elements}       
       
The popular Press-Schechter formalism (see, e.g., review         
in Loeb and Barkana, 2001) focuses main attention on the 
spherical collapse and the achieved final critical overdensity. 
It can be extended for elliptical clouds (see, e.g., Sheth 
and Tormen 2002). In spite of so strong restrictions on the 
process of collapse it successfully describes the simulated 
mass distribution of high density halos. In Zel'dovich theory 
of gravitational instability similar approach allows to find 
the mass function of all high 
density structure elements -- clouds, filaments, and walls or 
pancakes -- without any assumptions about their initial and 
final shapes, stability, relaxation and achieved overdensity. 
In this regard, our approach is much more general 
then the Press-Schechter formalism.        

The close connection of the observed objects with the 
initial power spectrum and the generic origin of galaxies 
and other observed elements of the LSS is clearly seen from 
the expression (\ref{d3w}) which applies to all structure 
elements. Due to 
symmetry of the relation (\ref{d3w}) with respect to variables 
$q_1,\,q_2\,\&\,q_3$ and because the mass functions are 
normalized, the restrictions (\ref{cp}, \ref{cf}\,\&\,\ref{ccl}), 
used for the determination of the mass fractions of different
structure elements, are now 
of no importance and they lead only to renumeration of coordinates. 
Thus, in the Zel'dovich approach, 
the mass functions of the various LSS elements differ only 
by their corresponding survival probability (see, e.g., Peacock 
\& Heavens 1990). 

\subsection{Initial shape of collapsed clouds}       

To find the mass function of structure elements we 
rewrite the general relation (\ref{d3w}) in spherical coordinates        
\[       
q_1=R\cos\theta,\quad q_2=R\sin\theta\cos\phi,\quad        
q_3=R\sin\theta\sin\phi\,,        
\]       
where $0\leq R\leq\infty$ and $0\leq\theta,\phi\leq\pi/2$ 
characterize the mass and shape of the collapsed structure 
elements in the Lagrangian coordinates:       
\be       
M\propto R^3,\quad \mbox{tg}\phi=q_3/q_2,\quad \mbox{tg}\theta=       
\sqrt{q_2^2+q_3^2}/q_1\,.        
\label{m3w}       
\ee       
For spherical clouds tg$\phi=$ 1, $\phi=\pi/4$, tg$\theta=
\sqrt{2}$. As was noted above, here we will neglect the  
restrictions (\ref{cp}, \ref{cf}\,\&\,\ref{ccl}) and 
differences in limits of $q_i$ (\ref{limits}). 
       
This result demonstrates that in the Zel'dovich approximation 
the distribution of shapes of initial clouds described by the 
parameters $\theta ~\&~\phi$ in (\ref{m3w}) is continuous 
and the finally achieved properties of structure elements, 
such as their shape, energy and overdensity, depend mainly 
upon the velocity or deformation field within clouds. In 
particular, even high density compact relaxed halos 
can be formed through the collapse of initially asymmetrical 
clouds. Some information about the initial shape of collapsed 
clouds is retained in the angular momentum of observed 
clouds discussed below.        
   
On the other hand, for all clouds, the first step of collapse 
is the formation of a pancake--like objects (Zel'dovich 1970; 
Shandarin et al. 1995) which are unstable and will rapidly break 
up into a system of low mass subclouds (Doroshkevich 
1980; Vishniac 1983). This instability stimulates  formation 
of numerous satellites of the largest central object and makes 
it difficult to observationally distinguish between isolated 
galaxies and these satellites.        
       
For the more massive clouds with $1> q_i\gg q_0$,        
$\eta_i\propto \sqrt{q_i}$, we have from (\ref{d3w}):        
\be       
d^3W = {\exp(-R\psi(\theta,\phi)/8\tau^2)\over        
\sqrt{R^3\cos\theta\sin 2\phi}}       
{dR^3 d\phi d\theta\over 6\pi^{3/2}\tau^3}\,,       
\label{rtp}       
\ee       
\[       
\psi(\theta,\phi)=cos\theta+\sqrt{2}\sin\theta\cos(\phi-\pi/4)\,.       
\]       
This relation shows that the fraction of initially massive spherical        
clouds with $\psi=\sqrt{3}$ is exponentially suppressed and asymmetric      
massive clouds dominate.        
     
For low mass clouds with $q_i< q_0$, we get for the mass function           
\[       
\eta_i\approx {1\over 2\tau}\sqrt{q_0\over 3}\left(1+       
{q_i^2\over 6q_0^2}+ ...\right)\,,       
\]       
\be       
d^3W\propto R^3dR^3\tau^{-3}\sin2\theta(1-\cos2\theta)\sin2\phi        
d\theta d\phi\,,       
\label{rtpm}       
\ee       
and the formation of low mass clouds with $M\leq M_\rho$ (\ref{mv})        
is suppressed. However, in this case the existence of approximately        
spherical initial clouds is much more probable.        

\subsection{Survival probability of objects}  
   
As it was shown above formation of low mass structure elements is usually suppressed 
because they could be absorbed  by larger objects formed at the same 
time and in the same region. This process is described by the survival 
probability of structure elements and it is different for clouds, filaments and walls. 
As was shown in DD99, for walls formed through the 1D compression 
along the $q_1$ axis the survival probability can be taken as 
\be
W_{surv}=\mbox{erf}(\sqrt{q_1/8\tau^2})\,.
\label{sw}
\ee
For filaments and clouds formed through the 2D and 3D compression 
the survival probabilities can be taken as 
\be
W_{surv}=\mbox{erf}(\sqrt{q_1/8\tau^2})\mbox{erf}(\sqrt{q_2/8\tau^2})\,,
\label{sf}
\ee
\be
W_{surv}=\mbox{erf}(\sqrt{q_1/8\tau^2})\mbox{erf}(\sqrt{q_2/8\tau^2})
\mbox{erf}(\sqrt{q_3/8\tau^2})\,,
\label{sc}
\ee
respectively. Evidently, so defined survival probability is small 
for low mass objects and $W_{surv}\rightarrow$ 1 for massive 
objects. 

Integration of the PDF (\ref{rtp}) corrected for the survival 
probability (\ref{sw}, \ref{sf} \& \ref{sc}), $W_{surv}d^3W$, over 
angular variables for given $R, q_0~\&~\tau$ allows one to find the 
mass functions of structure elements: clouds, filaments, walls and 
pancakes.  

\subsection{Mass function of structure elements}    
     
\begin{figure}     
\centering     
\epsfxsize=7.5 cm     
\epsfbox{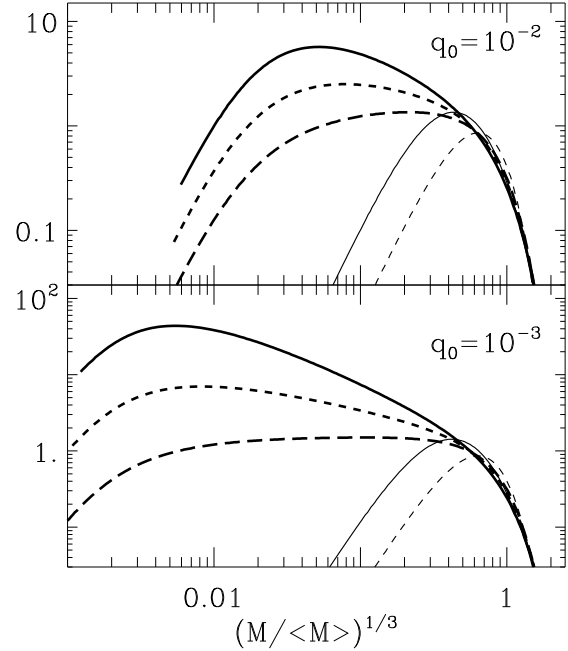}     
\vspace{1.2cm}     
\caption{Functions $N_w$ (thick solid line), $N_{f}$ (thick dashed line)
and $N_{c}$ (thick long dashed line) for two values of $q_0$ plotted 
vs. $(M/\langle M\rangle)^{1/3}$ for $\tau=0.3$. For $\tau\approx
\tau_r=\sqrt{q_0/6}$ and $\tau\approx 0.3\tau_r$, functions $N_w$
are plotted by thin solid and dashed lines, respectively.
}     
\label{fig3}
\end{figure}

\begin{figure}     
\centering     
\epsfxsize=7.5 cm     
\epsfbox{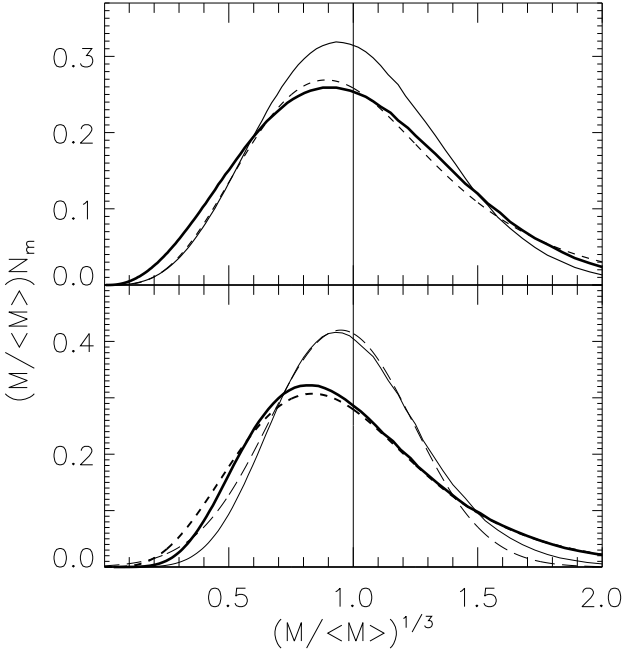}     
\vspace{1.2cm}     
\caption{
Top panel: Functions $M/\langle M\rangle~N_w$ (thick solid 
line) and $M/\langle M\rangle~N_c$ (thin solid line)  plotted vs.      
$(M/\langle M\rangle)^{1/3}$ for $\tau=0.3$. Fit (\ref{mnm03}) is 
drawn by dashed line. Bottom panel: functions $M/\langle M\rangle~N_w$ 
for $\tau\approx\tau_r=\sqrt{q_0/6}$ (thick solid line) and $\tau
\approx 0.3\tau_r$ (thin solid line) plotted vs. $(M/\langle M\rangle
)^{1/3}$. Fits (\ref{mnmrr}\,\&\,\ref{mnmr}) are drawn by thin dashed 
and long dashed lines, respectively.
} 
\label{fig4}    
\end{figure}

The mass function $N_m$ and the function $M/\langle M\rangle N_m$        
are plotted in Figs. \ref{fig3}~\&~\ref{fig4} for $q_0=10^{-2},
~\&~q_0=10^{-3}$ and for the three more interesting values of 
$\tau=0.3, \tau=\tau_r,~\&~\tau\approx 0.3\tau_r$. If $\tau\sim$ 
0.3 corresponds approximately to the present epoch (see Sec. 7.1), 
then $\tau\sim\tau_r$ (see (\ref{zr})) and $\tau\sim 0.3\tau_r$ 
describe the early period of structure formation, when the influence 
of small scale correlations of initial density and velocity fields 
is more important.       
       
As is seen from Fig. \ref{fig3} at $\tau\leq\tau_r$ the mass 
functions for all three kinds of structure elements are identical,
what is a direct consequence of strong correlations of small scale 
perturbations. However, the difference between these mass functions 
increases with time and at $\tau\approx$ 0.3 it becomes significant,
especially for $q_0=10^{-3}$.  
            
As is seen from comparison of Figs. \ref{fig3}~\&~\ref{fig4}, in 
all the cases the numerous low mass structure elements contain only 
negligible fraction of mass and for all $\tau$ the main mass is 
concentrated within structure elements with $M\sim (0.2 - 0.7)
\langle M\rangle$. For $\tau<~\tau_r$ the impact of the parameter 
$q_0$ on the shape of mass functions is negligible and, for both 
chosen values of $q_0$, the functions $M N_m(M)$ are well fitted 
by the Gauss function (\ref{mnmrr}). This result reflects again a 
strong small scale correlation of the initial density field and, 
as was discussed in Sec. 5.2, a small survival probability of 
low mass structure elements. 
      
At $\tau\geq\tau_r$ the mass function becomes wider 
because the continued formation of low mass structure elements 
is accompanied by progressive mass concentration within massive        
elements with $M\sim\langle M\rangle\gg M_\rho$ (\ref{mv}). As 
is seen in Fig. \ref{fig3}, during this period the small scale 
cutoff of initial power spectrum and the parameter $q_0$ provide 
the low mass cutoff of the mass function. For such redshifts, 
the mass functions for objects with  $M\leq\langle M\rangle$ 
and $M\geq\langle M\rangle$ are described respectively by power 
and exponential laws.
          
Disregarding the low mass 'tails' of mass functions, we         
fit the mass function for both values of $q_0$ by:       
\be       
x^3N_m(x)\approx 0.4\exp(-(x-0.95)^2/0.17),\quad        
\tau\approx 0.3\tau_r\,,       
\label{mnmrr}       
\ee       
\be       
x^3N_m(x)\approx 20x^2\exp(-4.1x){\rm erf}(x^2),
\quad \tau=\tau_r\,,       
\label{mnmr}       
\ee       
\be       
x^3N_m(x)\approx 12.5x^2\exp(-3.7x){\rm erf}(x^2),
\quad \tau=0.3,       
\label{mnm03}       
\ee       
where $x=(M/\langle M\rangle)^{1/3}$. These fits are plotted 
in Fig. \ref{fig3}\,. For $\tau\approx$ 1, we have        
\be       
x^3N_m(x)\approx 8. x^{3/2}\exp(-3.1x){\rm erf}       
(x^{9/4}) \,.       
\label{mnm1}       
\ee       
This function can be used to characterize the observed and      
simulated filaments and walls which are still forming.        
       
By definition, the mass functions must satisfy two 
normalization conditions 
\be       
3\int_0^\infty x^2N_m(x)dx\approx 3\int_0^\infty x^5N_m(x)dx\approx 1\,.       
\label{norm}       
\ee       
Fits (\ref{mnmrr} - \ref{mnm1}) better satisfy the second        
condition (\ref{norm}) while the first one is violated more        
strongly. These mass functions are similar to the expression        
(\ref{rtp}) and $N_m\propto \exp(-x)$, for $x\gg$ 1 and        
$\tau\geq \tau_r$.     
  
\subsection{Mean mass of objects}     
       
\begin{figure}       
\centering       
\epsfxsize=7.5 cm       
\epsfbox{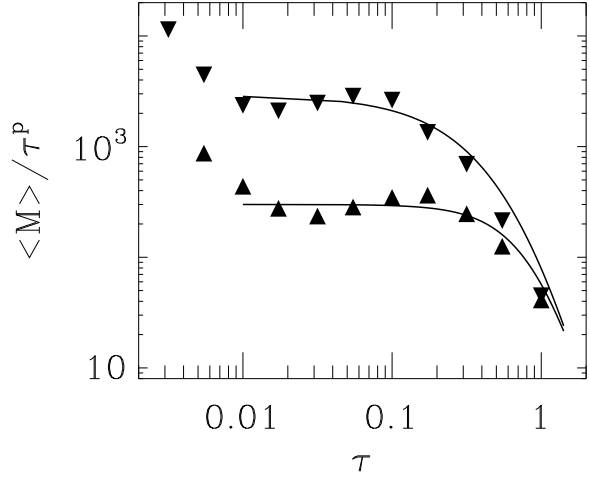}       
\vspace{1.2cm}       
\caption{Functions $\langle M\rangle/\tau^{4.3}$ (triangles, 
$q_0=10^{-2}$) and $\langle M\rangle/\tau^{5.2}$ (up down triangles,
 $q_0=10^{-3}$) plotted vs. $\tau$. Fits (\ref{mm2}\,\&\,\ref{mm3}) 
are drawn by solid lines, respectively.
} 
\label{fig5}      
\end{figure}

The mean mass of objects is quite sensitive to the coherent 
length of initial density field. For the same $q_0$ as above, 
the mean mass of clouds is plotted in Fig. \ref{fig5} versus 
$\tau$ together with fits    
\be       
\langle M\rangle \approx {3\cdot 10^2\tau^{4.3}\over        
(1+\tau^2)^{2.4}}M_v,\quad {\rm for}\quad  q_0=10^{-2}\,,       
\label{mm2}       
\ee       
\be       
\langle M\rangle\approx {3\cdot 10^3\tau^{5.2}\over        
(1+\tau^{1.2})^{5.2}}M_v,\quad {\rm for}\quad q_0=10^{-3}\,,        
\label{mm3}       
\ee       
which describe quite well the redshift evolution of $\langle 
M\rangle$ for $\tau\geq\tau_r$. For $\tau\leq\tau_r$ the much 
slower evolution of the mean mass is described by        
\be       
\langle M\rangle \approx \tau^3 M_v, \quad {\rm for}\quad 
q_0=10^{-2}\,,       
\label{mmt2}       
\ee       
\be       
\langle M\rangle\approx 4\cdot 10^{-2}\tau^3 M_v,\quad {\rm for}       
\quad q_0=10^{-3}\,,        
\label{mmt3}       
\ee       
what again emphasizes the impact of the small survival 
probability of clouds at $\tau\leq\tau_r$.        
        
For $\tau\geq\tau_r$, the mean masses of pancakes and filaments 
are less than those for  clouds by a factor of $\sim$ 
1.5 -- 2. 
       
\begin{figure*}     
\begin{minipage}{160mm}     
\centering     
\epsfxsize=15.cm     
\epsfbox{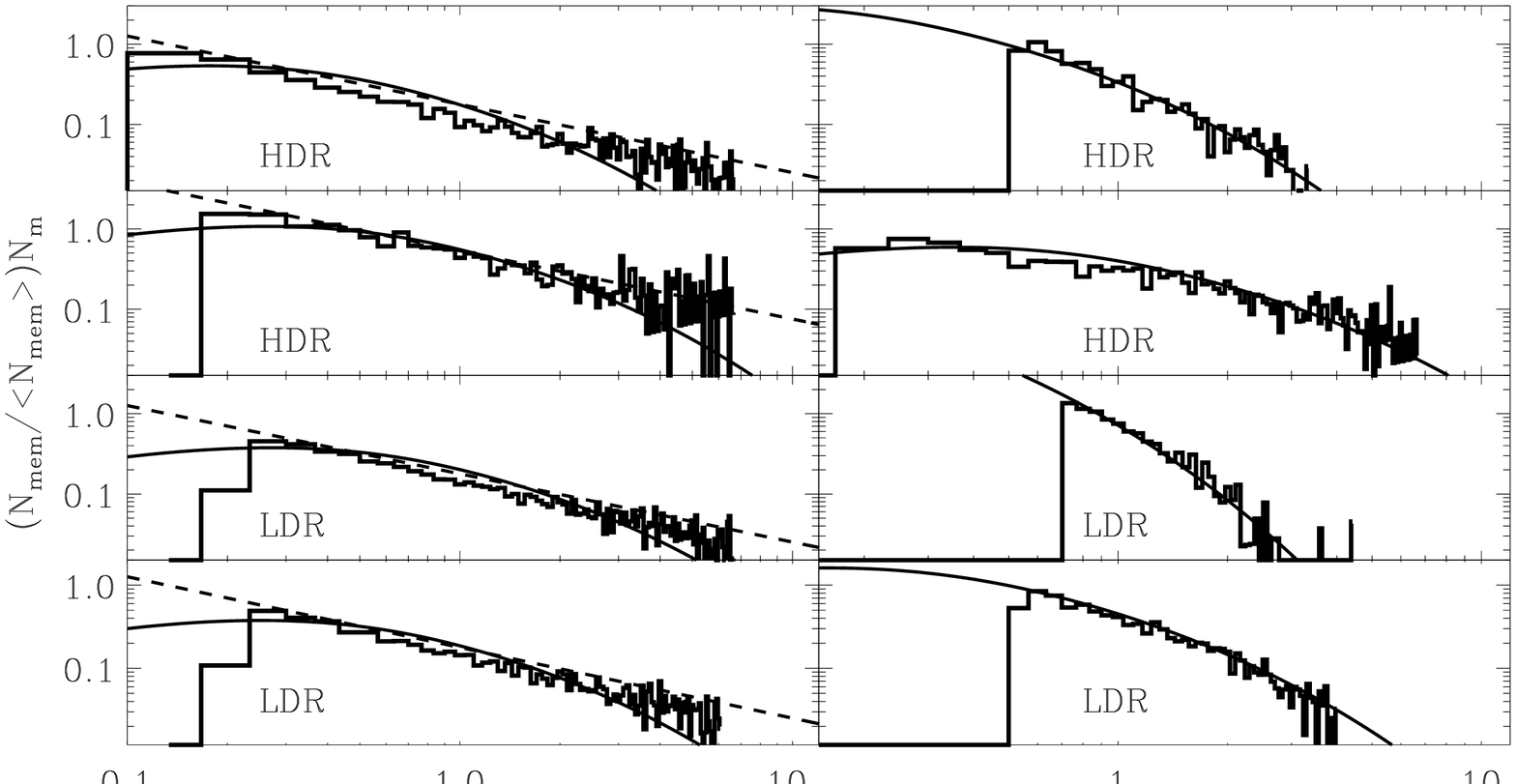}  
\vspace{1.2cm}     
\caption{Mass functions of galaxy clouds, $N_{mem}/\langle      
N_{mem}\rangle)N_m$, selected at redshifts $z=0$ (left panels) and      
 $z=4$ (right panels) in HDRs and LDRs for two threshold linking      
lengths. Fits (\ref{mnm03}) and 
(\ref{coag2}) are plotted by solid blue and green lines, respectively. 
  }     
\label{fig6}      
\end{minipage}     
\end{figure*}     
      
\subsection{Comparison with the Press -- Schechter 
formalism}     
       
The mass function $N_m$ describes all structure elements -- 
clouds, filaments, and walls or pancakes -- without assumptions 
about their shapes and achieved overdensity. However, the relation 
(\ref{mnm1}) is quite similar to the Press--Schechter mass function 
for scale--free power spectra and $k$ corresponding to typical
objects, $p\propto k^{-2}$:       
\be       
{M\over\langle M\rangle}N_{PS}d\xi = {8\over 45\sqrt{\pi}}
\xi^{1/6}\exp(-\xi^{1/3})~d\xi,				       
\label{ps}       
\ee       
\[       
\xi = 1.875~M/\langle M\rangle,\quad \langle \xi\rangle=15/8\,.       
\]       
       
In turn, for larger redshifts, $\tau\sim \tau_r$, and for a power        
spectrum with a cutoff at $k\sim k_{mx}$, both approaches predict        
the suppression of formation of low mass objects with $M\ll        
\langle M\rangle$ (Loeb\,\&\,Barkana 2001). This similarity is 
quite apparent as both        
relations are based on the same initial power spectrum. It indicates        
that the difference between the Press--Schechter and a more general        
Zel'dovich approach is quantitative rather then qualitative and        
these approaches are complementary to each other.        
   
\subsection{Impact of coagulation processes}  

All mass functions discussed above are related to the process of 
formation of structure elements and they do not take into account 
the later nonlinear evolution described by the coagulation equation 
(Smoluchowski 1916; Silk \& White 1978). This is not so important 
for walls and filaments for which the merging and coagulation 
are controlled mainly by the initial velocity field. But the non
linear evolution can 
essentially distort the mass function of high density clouds 
accumulated by richer walls and filaments.  

In contrast with expressions (\ref{mnmrr}-\ref{mnm1}) and 
(\ref{ps}) the coagulation process leads to the mass function 
\be
N_c(x)\propto x^{-\nu}\exp(-x),\quad x=M/\langle M\rangle\geq 
x_{min}\,,
\label{coag1}
\ee
where the power index $\nu\sim 3/2$ depends upon the aggregation 
rate (see, e.g., Silk \& White 1978). This mass function strongly 
differs from those discussed above. This means that the shape of 
observed and simulated mass functions measures the possible 
influence of coagulation processes. 
     
\subsection{Comparison with simulations}  

The theoretical fits (\ref{mnm03} \& \ref{ps}) can be compared      
with the mass functions found for matter distributions at redshifts 
$z=4$, and $z=0$ in the high resolution simulation (Klypin et al. 
1999; Schmalzing et al. 1999) discussed in Sec. 4.1.
As was described in Sec. 4.1, at both redshifts the full samples 
were divided into subsamples of high and low density regions (HDRs 
and LDRs) dominated by walls and filaments.  
    
The mass functions of high density clouds found for the HDRs and      
LDRs separately are plotted in Fig. \ref{fig6}. The basic parameters 
of the same samples of selected clouds are listed in Table 1, where 
$\delta_{thr}$ and $\langle \delta\rangle$ are threshold and mean 
overdensities of clouds above the mean density of the sample, 
$f_{pnt}$ is the fraction of points accumulated by clouds, $N_{cl}$ 
and $\langle N_{mem}\rangle$ are the number and the mean richness 
of clouds. At both redshifts, very massive structure elements are      
formed through the percolation process and they cannot be described      
by the expressions (\ref{mnm03} \& \ref{ps}). For this reason, they      
were excluded from the analysis of the mass distribution. However,      
these clouds are included in estimates of the matter fraction      
$f_{pnt}$ accumulated by structure elements as they present actual
filaments and walls. The difference 
between both $f_{pnt}$ and $\langle N_{mem}\rangle$ for clouds 
selected with the same $\delta_{thr}$ within HDRs and LDRs confirms 
a significant impact of environment on the properties of clouds.     

At both redshifts, the samples selected with high $\delta_{thr}$ 
represent properties of small fraction of high density clouds, 
while samples selected with small $\delta_{thr}$ are formed mainly 
by unrelaxed filaments and walls.         
The cutoff of the simulated mass functions at low masses caused      
by a finite resolution increases the measured mean richness of      
selected clouds as compared with theoretical expectations (\ref{mm2}).      
Because of this, the mean richness was used as a fit parameter in      
the relations (\ref{mnm03}) plotted in Fig. \ref{fig6}. However, 
the shape of the mass function (\ref{mnm03}) was not altered. As 
is seen from Fig. \ref{fig6}, this expression fits well the 
simulated mass functions for both high density clouds and only 
partly relaxed filaments and walls.      
      
\begin{table}     
\caption{Parameters of structure elements selected in HDRs and LDRs      
at $z$=0 \& 4.}      
\begin{center}     
\begin{tabular}{l lrl rc}      
\hline     
&$\delta_{thr}$&$\langle \delta\rangle$&$f_{pnt}$&$N_{cl}$&$\langle      
N_{mem}\rangle$\\      
\hline     
&&&$z=4$&&\\     
HDR&490    &3 007&0.015&2 187&101\\     
HDR&~~~1.6&15&0.37 & 5 311&313\\     
LDR&~~~7.1&139&0.004&  963&~~75\\     
LDR&~~~1.6& 15&0.036&6 104&108\\     
\hline     
&&$z=0$&&\\     
HDR&86  &1 200&0.18&11 155&273\\     
HDR&22  &  645&0.33& 2 908&342\\     
LDR&22  &  712&0.21& 8 484&270\\     
LDR&~~1.&   62&0.43& 9 629&261\\     
\end{tabular}     
     
The few richest clouds formed through percolation process are 
excluded from calculations of $\langle N_{mem}\rangle$.      
\end{center}     
\end{table} 

However, at the redshift $z=0$ the simulated mass functions are well 
fitted also by a power law 
\be
xN_m(x)\propto x^{-0.85},\quad x=N_{mem}/\langle N_{mem}\rangle\,,
\label{coag2}
\ee
similar to (\ref{coag1}). It indicates the possible  
influence of the coagulation processes on the parameters of 
selected high density clouds and even filaments and walls. At 
the redshift $z=4$ this influence is weak and the mass functions 
are evidently exponential.

It is especially important that at both redshifts the relations 
(\ref{mnm03} \& \ref{ps}) successfully reproduce the mass functions 
of unrelaxed filaments and walls which are far from the spherical 
shape. This fact demonstrates a moderate influence of the shape 
of collapsed clouds on their mass and the validity of Zel'dovich 
approach which considers the dynamical characteristics of collapsed 
clouds rather than their shape.       

\subsection{The angular momentum of collapsed clouds}       
       
Using the statistical approach it is possible to consider the angular        
momentum of the collapsed clouds. For this purpose we  use the 
general equation (\ref{eq1}) together with the corresponding 
expression for the velocity of fluid element        
\be       
v_i={dr_i\over dt} = H(z)r_i-{H(z)\over 1+z}(\beta(z)-1)B(z)S_i\,,       
\label{v1}       
\ee       
where $\beta(z)=(1+z)d\ln{B}/dz$, $H(z)$ is the Hubble parameter, and 
$B(z)$ was introduced by (\ref{B1}). As usual, we 
define the angular momentum of a particle as        
\be       
j_i=\epsilon_{ijk}r_jv_k = J_0\epsilon_{ijk}{\tilde q}_jS_k,\quad 
J_0={H(z)(\beta-1)B(z)\over (1+z)^{2}}       
\label{mm1}       
\ee       
where $\epsilon_{ijk}$ is the unit antisymmetric tensor and the 
function $J_0(z)$ describes time variations of the angular momentum. 
Let us note that the description of the angular momentum through 
the deformation tensor (see, e.g., White 1984) is useful 
methodically but cannot be applied to more massive objects because 
of the small correlation scale of this tensor.       
       
The angular momentum of a cloud is defined by the integral over 
the corresponding collapsed volume, $V$,       
\be       
\langle J^2\rangle = J_0^2{\sigma_s^2l_v^2\over 3V^2}\int_Vd^3       
\tilde{p}d^3\tilde{q}~I({\bf p},{\bf q})\,,       
\label{mm22}       
\ee       
\[       
I({\bf p},{\bf q})=2({\bf pq})G_1(|{\bf p}-{\bf q}|)       
+(p^2q^2-({\bf pq})^2)G_2(|{\bf p}-{\bf q}|)\,,       
\]       
(the functions $G_1 ~\&~G_2$ are introduced in DD99 and Appendix A).       
It depends upon the size and shape of the collapsed region,
statistical characteristics of which are described by (\ref{rtp}). 
For low mass early formed clouds with $p\leq q_0$, $q\leq q_0$, we 
get        
\be       
I\approx [3({\bf pq})^2-p^2q^2]/q_0\,,       
\label{qlq}       
\ee       
and, for example, for ellipsoidal clouds with axes $a_1, a_2, ~\&~a_3$        
we have        
\be       
\int_Vd^3pd^3qI\propto{1\over q_0}[a_{12}^2+a_{13}^2+       
a_{23}^2],\quad a_{ij}=a_i^2-a_j^2\,,       
\label{ell}       
\ee       
what is identical with the expression found already in Doroshkevich 
(1970).         
     
For larger clouds with $p\gg q_0$, $q\gg q_0$, we get        
\be       
I\approx {5({\bf pq})^2-p^2q^2-2({\bf pq})(p^2+q^2)\over
\sqrt{p^2+q^2-2({\bf pq})}}\,,       
\label{qbq}       
\ee       
and for spherical clouds $\langle J^2\rangle=0$. Numerical 
integration of (55) shows that for large elliptical clouds 
$\langle J^2\rangle$ depends on the axes $a_i$ in a similar 
way as for small clouds (\ref{ell}).        
   
\section{Statistical characteristics of filaments}     

\begin{figure}
\centering
\vspace{0.2cm}
\epsfxsize=7cm
\epsfbox{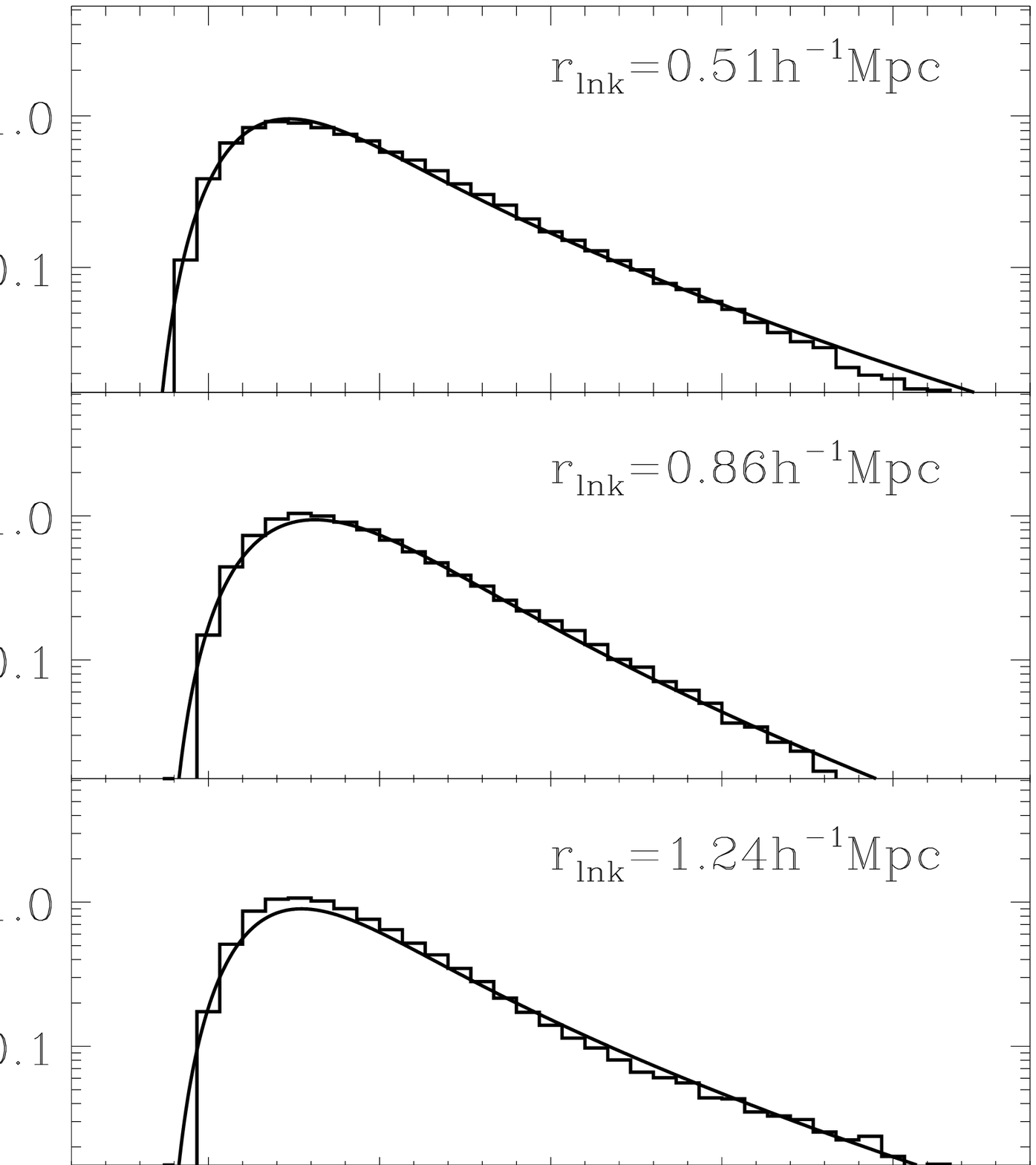}
\vspace{1.2cm}
\caption{Distribution function, $N_{fil}$, for the linear 
density of DM particles in filaments selected at three 
linking lengths, $r_{lnk}$. Fits (\ref{fitfil}) are plotted 
by solid lines. 
} 
\label{fig7}      
\end{figure}

In addition to the fraction of matter accumulated by filaments 
(\ref{wf}) and the mass function of filaments given by (\ref{mnmrr}-
\ref{mnm1}), we will consider here also the linear density of 
galaxies along a filament, $\Sigma_{fil}$, defined as a mass per 
unit length of filament, and the mean surface density of filaments, 
$\sigma_{fil}$, defined as the mean number of filaments intersecting 
a unit area of arbitrary orientation. These characteristics 
depend upon the threshold linking length, $r_{lnk}$, used for 
the filament selection, which determines the threshold overdensity 
bounding the filaments. Now only richer filaments can be selected 
in both observed and simulated catalogues what restricts its 
quantitative characteristics. Because of this, here we will  
only discuss characteristics of richer filaments. 

The distribution function of filaments linear density describes 
their frequency distribution with respect to the amount of matter 
per unit length of filaments. As was discussed in DD99, this 
function can be obtained by integrating  (\ref{d3w}) over all $q_3$ 
and over the ratio $q_1/q_2$. For $q_0\ll 1, q_0\ll q\leq 1$ and 
using the survival probability for low mass filaments given by 
(\ref{sf}) we have for the PDF of richer filaments with 
$\Sigma_{fil}\propto q_1^2+q_2^2\ge\, <\Sigma_{fil}>$ 
\be
N_{fil}\approx {1.5\over\langle\Sigma_{fil}\rangle}\exp(
-\sqrt{3\Sigma_{fil}/\langle\Sigma_{fil}\rangle}),\quad 
\langle\Sigma_{fil}\rangle\approx 48\tau^4.
\label{nfil}
\ee

In Fig. \ref{fig7} the PDFs, $N_{fil}$, are plotted for filaments 
selected from two simulated DM distributions at $z=0$ (Jenkins 
et al. 1998, top panel, Cole et al. 1998, middle and bottom 
panels) for three linking lengths $r_{lnk}$. The linear density 
of matter, $\Sigma_{fil}$, was measured by the ratio of the number 
of points and the length of the minimal spanning tree for each 
filament of the sample. 

The PDF, $N_{fil}$, is well fitted by 
\be
N_{fil}=a_0{\rm erf}^4[a_1(x-x_0)]\exp(-\sqrt{a_2(x-x_0)})\,, 
\label{fitfil}
\ee
where $x=\Sigma_{fil}/\langle\Sigma_{fil}\rangle$ and
\[
a_0=50,\quad ~a_1=2.5,\quad  a_2=27,\quad  x_0=0.3\,, 
\]
\[a_0=180,\quad a_1=2,~~\quad a_2=42,\quad x_0=0.35\,,
\]
\[
a_0=60,\quad ~a_1=2.5,\quad a_2=31,\quad x_0=0.35\,, 
\]
for these samples, respectively. Here the cut--off of the PDF at 
$x=x_0$ reflects the limited resolution with which the filaments 
were selected. This cut--off increases $\langle\Sigma_{fil}\rangle$ 
and changes the parameters $a_0, a_1, a_2$ in (\ref{fitfil}) with 
respect to the expected ones (\ref{nfil}). 
The measured mean linear density, $\langle\Sigma_{fil}\rangle$, 
depends upon the linking length used for the selection of 
filaments, and for the samples used we have 
\[
r_{lnk}\Sigma_{fil}=3.4,\quad r_{lnk}\Sigma_{fil}=3.1,\quad 
r_{lnk}\Sigma_{fil}=3.1\,,
\] 
respectively.
At the same time, for all samples at $\Sigma_{fil}\geq\langle
\Sigma_{fil}\rangle$ the exponential term dominates what is  
consistent with theoretical expectations (\ref{nfil}). 

For the surface density of filaments, $\langle\sigma_{fil}\rangle$, 
and for their mean separation, $\langle D_{fil}\rangle=\langle
\sigma_{fil}\rangle^{-1/2}$ we get, respectively:
\be
\langle\sigma_{fil}\rangle\approx {W_f(r_{lnk})\langle n\rangle
\over (1+z)^2\langle\Sigma_{fil}\rangle},~ \langle D_{fil}
\rangle=\sqrt{(1+z)^2\langle\Sigma_{fil}\rangle\over W_f(r_{lnk})
\langle n\rangle}.
\label{sigf}
\ee 
Here $W_f(r_{lnk})$ and $\langle n\rangle$ are the fraction of 
objects accumulated by filaments selected with a given $r_{lnk}$ 
and the mean density of objects in the sample. Evidently, $W_f\ll 
1$ for $r_{lnk}\langle n\rangle^{1/3}\leq 1$ and $W_f\rightarrow 
0.3 - 0.5$ for larger $r_{lnk}$. The variations of $W_f$ strongly 
influence $\langle\sigma_{fil}\rangle$ and $\langle D_{fil}
\rangle$ and determine how they vary with $r_{lnk}$, which is 
strongly connected with the
overdensity of selected filaments. However, these estimates of 
$\langle\sigma_{fil}\rangle$ do not take into account possible 
accumulation of filaments by walls and therefore they should be 
corrected by comparing them with observations and simulations.
 
\section{Characteristics of walls and pancakes}     
     
From the PDF of the differences of displacements (\ref{ws}) it is     
possible to extract important approximate characteristics      
of walls and less massive pancakes which can be directly compared      
with available observations. Some of them were introduced in DD99      
and successfully compared with simulated and observed 
characteristics of walls. Other characteristics discussed in this 
Section are successfully used for description of low mass
pancakes observed as Ly-$\alpha$ absorbers at large redshifts  
(Demia\'nski, Doroshkevich \& Turchaninov 2003) and for the 
determination of the initial power spectrum at small scale 
(Demia\'nski and Doroshkevich 2003).
     
\subsection{Distribution function of wall sizes}     
     
The distribution function of sizes of walls     
describes their frequency distribution with respect to their 
Lagrangian size what is identical to their surface density, 
$m_{w}$, defined as the mass per unit surface area of the wall 
at the moment of its formation. This distribution function is 
obtained by integrating (\ref{d3w}) over all $q_2$ and $q_3$ 
for $q_0\ll 1$, $q\leq 1$ and it is given by: 
\be     
N_{w}\approx {1\over \sqrt{2\pi}\tau}{\Theta_w(q_0/q_w)
\over\sqrt{q_{w}}}\exp\left(-{q_{w}\over 8\tau^2}\right)
\mbox{erf}\left(\sqrt{q_{w}\over 8\tau^2}\right),     
\label{ww}     
\ee     
\[     
q_{w} = {m_{w}\over l_v\langle \rho\rangle}={|{\bf q}_1-     
{\bf q}_2|\over l_v},\quad \int_0^\infty N_{w}(q_{w})~dq_{w} = 1,     
\]     
\[     
\langle q_{w}\rangle = \int_0^\infty q_{w}     
N_{w}(q_{w})~dq_{w}\approx 8(0.5+1/\pi)\tau^2\approx 6.55\tau^2\,,     
\]     
\[     
\Theta_w(y)=4\sqrt{q_w}d\mu(q_w)/dq_w\,.     
\]  
The factors erf$(\sqrt{q_{w}/ 8\tau^2})$ and $\Theta_w$ 
in (\ref{ww}) describe the survival probability of walls and the      
influence of the coherent length of initial density field, $q_0$. 
For $q_w\gg q_0$, $\Theta_w$=1 and $N_w$ becomes identical to the one 
found in DD99.     
     
\subsection{Transverse characteristics of walls and pancakes}     
     
For some applications we have to estimate the transverse      
characteristics of pancakes such as the distribution function      
of Lagrangian size and the mean real size of pancakes. These      
characteristics can be found with the method used above.      
     
For the frequency distribution of the walls with Lagrangian      
transverse sizes, $q_2~\&~q_3$, we get from (\ref{mu1}, \ref{mu2},     
\ref{d3w})      
\be     
N(\eta_2)d\eta_2 = {2\over\sqrt{\pi}}\exp(-\eta_2^2)d\eta_2,\quad      
\eta_0\leq \eta_2\leq\infty\,,     
\label{tr}     
\ee     
\[     
\langle\eta_2\rangle = {1/\sqrt{\pi}},\quad      
\langle\eta_2^2\rangle = {\langle q_2\rangle/ 8\tau^2} = 1/2\,,     
\]     
and the same distribution for $\eta_3 ~\&~q_3$. In these cases,     
for the low mass objects the merging and percolation are not so    
important and the functions (\ref{tr}) are not corrected for the    
survival probability.      
     
The mean transverse size of expanded, $\Delta r_e$, and compressed,      
$\Delta r_c$, pancakes can be found from relations (\ref{eq2},      
\ref{ws}) and (\ref{tr}), we have      
\be     
\langle\Delta r_e\rangle ={\Delta_0\over 2}\left(1+{1\over\pi}
\right),\quad\langle\Delta r_e^2\rangle = {7\Delta_0^2\over 8}
\left(1+{8\over 7\pi}\right)\,,     
\label{dr}     
\ee     
\[     
\langle\Delta r_c\rangle\approx {\Delta_0\over 2}\left(1+{1\over 
2\pi}\right),\quad      
\langle\Delta r_c^2\rangle\approx{13\Delta_0^2\over 16}
\left(1+{10\over 13\pi}\right)\,,      
\]     
\[     
\Delta_0={8\tau^2\over 1+z}l_v\,.     
\]     
At small redshifts for the $\Lambda$CDM cosmological model with      
$\Gamma=0.2$, $8\tau_0^2\approx 0.5$, the expected mean transverse      
size of walls $\sim 0.5l_v\approx 16h^{-1}$Mpc is similar to the      
observed one (Doroshkevich et al. 2001).      
     
\subsection{Distribution function of pancakes surface density}     
     
After pancake formation, the transverse compression and/or 
expansion of matter changes its surface density and other 
characteristics. However, the direct analysis indicates that 
the PDF of pancakes surface density given by (\ref{ww}) 
only weakly depends upon these deformations. 
     
The evolution of the pancake surface area, $S_{pan}$, is described       
by the relation (\ref{eq2}) as follows:     
\[     
S_{pan}(z)\propto {\Delta r_2\Delta r_3\over q_2q_3} =      
{1\over (1+z)^2}\left(1-{s_2\over\sqrt{2}\eta_2}\right)     
\left(1-{s_3\over\sqrt{2}\eta_3}\right)     
\]     
Therefore, for a pancake formed at a redshift $z_f$ with the
surface density      
\[     
\sigma_{pan}(z_f) = l_v(1+z_f)^2 q_1\,,     
\]     
the surface density at a redshift $z$ is     
\be     
\sigma_{pan}(z)=\sigma_{pan}(z_f)S_{pan}(z_f)/S_{pan}(z)\,.     
\label{sp}     
\ee     
This means that the fraction of matter accumulated by pancakes with      
the surface density $\geq \sigma_{pan}$ depends upon their      
transversal sizes, $q_2~\&~q_3$, and instead of (\ref{wp}) we get 
from (\ref{ws}):     
\be     
W_\sigma = {3\over\pi}\int_{-\infty}^{\eta_2,\eta_3}     
dx_2dx_3\exp(-x_2^2-x_3^2) \times      
\label{wwp}     
\ee     
\[     
\mbox{\rm erfc}\left(-\eta_p^2{[1-x_2/\eta_2(z)][1-x_3/\eta_3(z)]     
\over[1-x_2/\eta_2(z_f)][1-x_3/\eta_3(z_f)]}\right)\,.     
\] 
As before, here $d^3W_\sigma/d^3q_i$ is the PDF similar to 
(\ref{d3w}) and, after integration of this PDF over all $q_2, 
q_3$, we obtain (\ref{wwp}) with $q_2,q_3\rightarrow
\infty$. So, we get       
\[     
N_\sigma\rightarrow \exp(-\eta_p^2){d\eta_p\over d\sigma_{pan}},
\quad \eta_p^2 = {\sigma_{pan}\over 8\tau^2(1+z)^2}\,.     
\]     
     
If the survival probability of a pancake is erf($\eta_p$) 
then the normalized PDF of  surface 
density is given by  (\ref{ww}), where $q_w$ and $\tau$ are 
replaced by $\sigma_{pan}$ and $\tau_z=(1+z)\tau$, and, for 
example,       
\be     
\langle \sigma_{pan}\rangle = 8(0.5+1/\pi)(1+z)^2\tau^2\,.     
\label{qpan}     
\ee     
However, if we retain the expression for the survival probability      
used in (\ref{ww}) then the statistical characteristics of pancakes      
will depend upon both the redshift of pancake's formation and      
its current redshift.      
     
These results indicate that at high redshifts, when $(1+z)\tau(z)     
\approx {\rm const}.$ the expansion and compression of pancakes      
approximately compensate each other and the PDF (62) or (66), the 
mean surface density, $\langle \sigma_{pan}\rangle$, and other 
average characteristics of pancakes only weakly depend upon the 
redshift. They indicate also that, in spite of the strong evolution 
of each individual pancake, the statistical description (\ref{ww}) 
remains valid also when we consider each wall and each pancake as 
formed at their current redshifts. 

Application of these results to absorbers observed in a wide 
range of redshifts (Demia\'nski, Doroshkevich \& Turchaninov 2003) 
confirms these conclusions.
     
\subsection{Mean comoving linear number density of walls}     
     
Using the relations (\ref{d3w}~\&~\ref{ww}) it is also possible to 
obtain an {\it approximate} estimate of the mean comoving linear 
number density of recently formed walls, that is the mean number 
of walls per unit distance along a straight line. For richer walls      
with a threshold surface density $q_{thr}\gg q_0$ the small scale      
fluctuations of density are not important and this function can be      
written as follows:     
\be     
\langle n_w(\geq q_{thr})\rangle \approx {3\over 8}{\mbox{erfc}
(\eta_{thr})\over l_v}{(1+z)^2 \over \langle q_w(q_{thr})\rangle}\,,     
\label{nnw}     
\ee     
\[     
\langle q_w\rangle=4\tau^2\left[1+     
{4\sqrt{\pi}\eta_{thr}~\mbox{erf}(\eta_{thr})+2\exp(-\eta_{thr}^2)     
\over \pi\exp(\eta_{thr}^2)\mbox{erfc}(\eta_{thr})     
[1+\mbox{erf}(\eta_{thr})]}\right]\,,     
\]     
where $\eta_{thr}^2=q_{thr}/8\tau^2$, and the factor $(1+z)^2$ 
describes the expansion of the universe. For $q_{thr}\rightarrow 0,
~\eta_{thr}\ll$ 1, we have $\langle q_w(q_{thr})\rangle\rightarrow
\langle q_w(0)\rangle$ as given by (\ref{ww}) and the mean linear 
number density of pancakes increases as      
\[     
\langle n_w(\geq q_{thr})\rangle\propto (1+z)^2 \tau^{-2}\propto      
(1+z)^4\,.      
\]     
For $\eta_{thr}\gg$ 1 we have $\langle q_w(q_{thr})\rangle\approx      
q_{thr}$. Similar relations can also be written for the threshold      
surface density of pancakes.   Application of these results to the 
Lyman-$\alpha$ clouds observed in a wide range of redshifts 
(Demia\'nski, Doroshkevich \& Turchaninov 2003; Demia\'nski, 
Doroshkevich  2003) confirms that this relation correctly describes 
the observed redshift distribution of stronger lines in the 
Ly-$\alpha$ forest. 
     
\subsection{Coagulation approach}  

To describe the nonlinear evolution of walls observed in deep galaxy
surveys we can also use  
the 1D version of the coagulation equation (Smoluchowski 1916; Silk 
\& White 1978) which can be written in the comoving space as follows:
\be
{\partial n(q,\tau)\over \partial \tau} = {1\over 2}\int_0^q dx 
P(x,q-x,\tau)n(x,\tau)n(q-x,\tau)-
\label{smol1}
\ee
\[
n(q,\tau)\int_0^\infty dx P(x,q,\tau)n(x,\tau)\,.
\]
Here $n(q,\tau)$ is the comoving linear 
number density of walls with the dimensionless surface density $q$, 
$P(x,q,\tau)$ is the rate of aggregation of walls, and it is assumed 
that walls accumulate the main fraction of mass. Using this approach 
it is possible to find the linear number density of walls but their 
evolution depends on the unknown aggregation rate $P(x,q,
\tau)$ which is a complicated function of $q~\&~\tau$. In particular, 
it depends upon the initial power spectrum. 

The simplest reasonable solution of the coagulation equation, similar 
to (\ref{ww}), can be written as follows:
\be
n(q,\tau)={16\over l_vP_0^2\tau^4}\exp\left(-{4q\over 
P_0\tau^2}\right),\quad P(x,q,\tau)=P_0\tau\,,
\label{smol2}
\ee
and for the mean linear number density of walls we get:
\be
\langle n(q_{thr},\tau)\rangle = (1+z)^2\int_{q_{thr}}^\infty dx 
~n(x,\tau) 
\label{smol3}
\ee
\[
= {4(1+z)^2\over l_vP_0\tau^2}\exp\left(-{4q_{thr}\over P_0
\tau^2}\right)\,,
\]
what is similar to (\ref{nnw}). For more complicated aggregation 
rates $P(x,q,\tau)$ some solutions were given by Silk \& White 
(1978). 

\subsection{Linear number density of low mass pancakes}     
     
The approach discussed in Sec. 7.4 neglects the influence of small 
scale perturbations and approximately characterizes only the mean      
linear number density of richer walls. For low mass pancakes of a       
size comparable with the coherent scale of initial density field,      
$q_w\sim q_0$, the mean linear number density of pancakes with a      
threshold size $q_{thr}$ depends upon the spectral moment, $m_0$, 
and $q_0$ (\ref{lv}). It can be found with the standard technique 
(see, e.g., BBKS) used to describe the condition that a random 
function exceeds a certain value, we get:
\be     
\langle n(\geq q_{thr})\rangle\approx {\sqrt{3}(1+z)^2\Theta(y)     
\over 16\pi l_v\tau(z)\sqrt{q_0}}     
\Phi(\eta_{thr},\eta_2,\eta_3)\,,     
\label{nq1}     
\ee     
\[     
\Phi= {\mbox{erf}(\eta_{thr})\over\eta_{thr}}\int_{-\infty}^{\eta_2,     
\eta_3}{dx_2dx_3\over \pi} ~exp(-\eta_{thr}^2-x_2^2-x_3^2)\,,     
\]     
\[     
\Theta^2(y)= 1+{q_0\over 3}{d^2\xi_v\over d^2q}\approx      
{y^2\over p^2}\left[1+{4+(p-1)^2\over p(1+p)^2}+{1\over p^2(1+p)}     
\right]     
\]     
\[     
y=q_{thr}/q_0,\quad p=\sqrt{1+y^2},\quad\eta_{thr}^2=q_{thr}/8     
\tau^2(z)\,.     
\]      
Here the factors $(1+z)^2$ and $\mbox{erf}(\eta_{thr})$ describe 
the impact of expansion of the universe and merging of pancakes. 
The factor $\Theta(y)$ in (\ref{nq1}) introduces corrections for 
the case $q\leq q_0$, while $\Theta\rightarrow$ 1, for $q\gg q_0$,      
$y\gg1$. The density (\ref{nq1}) depends upon transverse motions 
characterized by the parameters $\eta_2, \eta_3$. For $\eta_2, 
\eta_3\gg$ 1, $\eta_{thr}\ll$ 1, $\langle n(\geq q_{thr})\rangle
\propto (1+z)^3$ and it grows not so fast as for the richer walls 
(\ref{nnw}) because formation of pancakes with $q\leq q_0$ is 
suppressed.     
     
The relation (\ref{nq1}) characterizes pancakes by their threshold 
size at the redshift of formation and neglects evolution of pancakes      
after they are formed. However, the surface density of formed pancakes      
is changing because of their transversal compression and/or expansion 
which shifts some of the pancakes under and/or over the observational 
threshold. This problem is quite similar to that discussed in Sec. 
7.3, where it was noticed that the surface density is a more adequate 
characteristic of DM pancakes which takes into  account these 
variations.
     
In the relations (\ref{nq1}) the threshold size of pancakes,
$q_{thr}$, together with their transverse sizes appears only in the 
function $\Phi(\eta_{thr},\eta_2,\eta_3)$. This means that to go 
from the threshold size to the threshold surface density we must 
link $q_{thr}$ and $\sigma_{thr}$ with the expression (\ref{sp})      
and find a new function $\Phi(\sigma_{thr},\eta_2,\eta_3)$.       
This procedure is quite similar to that used in Sec. 7.3,      
and, for $\eta_2\rightarrow\infty$, $\eta_3\rightarrow\infty$,      
we get instead of (\ref{nq1}) that     
\be     
\langle n(\geq \sigma_{thr})\rangle\approx {\sqrt{3}(1+z)^2\Theta\over      
16\pi l_v\tau(z)\sqrt{q_0}}\exp(-\eta_s^2){\mbox{erf}(\eta_s)\over     
\eta_s}\,,     
\label{nabs}     
\ee     
\[     
\eta_s^2 = {\sigma_{thr}\over 8\tau^2(1+z)^2}\,.     
\]     
This result demonstrates again that, as was discussed in Sec. 7.3,      
the transverse compression and/or expansion of pancakes compensate 
each other and it does not change their statistical characteristics, 
if we consider each pancake as formed at its current redshift.      
     
Both expressions (\ref{nq1}) and (\ref{nabs}) were used in 
Demia\'nski, Doroshkevich \& Turchaninov (2003) to describe the 
observed evolution of the mean linear number density of pancakes 
and to estimate the important parameter $q_0$ and the moment $m_0$ 
of the initial power spectrum. 
    
\section{Summary and discussion}       

In this paper we continue the statistical description of the 
process of LSS formation and evolution based on the Zel'dovich 
theory of nonlinear gravitational instability. First results 
obtained in DD99, DDMT and Demia\'nski et al. (2003) show a 
significant potential of this approach. Here we are allowing for 
deformation of pancakes after their collapse along the axis 
of the most rapid compression, interaction of large and small 
scale perturbations and the impact of small scale cutoff in 
the initial power spectrum. This extension allows one to consider 
three important problems.

First of all, we are able to find the mass functions and fractions 
of matter accumulated by the LSS elements, namely, pancakes, 
filaments and halos for a  wide range of redshifts. As was shown 
in Sec. 6, the mass functions describe reasonably well simulated 
mass distributions at all redshifts that emphasizes the 
generic character of the processes of formation of all structure 
elements. These functions provide quantitative description 
of the LSS evolution that in itself is an important problem. 

Secondly, we discuss the interaction of large and small scale 
perturbations, which  manifests itself as a strong 
concentration of galaxies within filaments and walls 
observed at small redshifts. This interaction is complex 
and it requires more detailed investigation.

Thirdly, we derived the mass function and the mean linear 
number density of pancakes at high redshifts. Both functions 
play an important role in the interpretation of the Ly-$\alpha$ 
forest observed in spectra of the farthest quasars. Results 
obtained in Sec. 7 are successfully applied in Demia\'nski et 
al. (2003) and Demia\'nski \& Doroshkevich (2003) for detailed 
description of observed absorbers and, in particular, lead to 
estimates of the spectral moment $m_0$ and the mass of the 
dominant fraction of dark matter particles. 
        
The rapid growth of the observed concentration of neutral        
hydrogen at redshifts $z\sim$ 6 (Djorgovski et al. 2001;        
Becker et al. 2001; Pentericci et al. 2001; Fan et al. 2001)        
is an evidence in favor of the reionization of the Universe at        
this redshift. These observations stimulate discussions        
of the reheating of the universe and, in particular, of the 
warm dark matter (WDM) models (see, e.g., Barkana et al. 2001;        
Loeb \& Barkana 2001). This means that        
it is worthwhile to find direct estimates of the small scale 
initial power spectrum        
and its influence on the LSS formation.
Results obtained in Secs. 4, 5\,\&\,7 are quite important for 
such investigations. 

\subsection{Amplitude of perturbations}       
       
In the Zel'dovich theory the evolution of structure can be 
suitably characterized by an effective dimensionless `time', 
$\tau(z,\Omega_m,h)$, introduced by (\ref{tau}):       
\be       
\tau(z) = \tau_0(\Omega_m,h)B(\Omega_m,z),\quad \tau_0 =        
{\sigma_{s}\over\sqrt{3}l_v}= {A\over \sqrt{3}}m_{-2}^{3/2}\,,		       
\label{tauu}       
\ee       
which describes the evolutionary stage achieved in the model. 
The function $B(z)$, the amplitude $A$ and the spectral moment 
$m_{-2}$ were introduced in (\ref{B1}--\ref{lv}).       
       
$\tau_0$ can be measured by different methods. More popular       
characteristics of the amplitude are $\sigma_8$ -- the variance       
of mass in a randomly placed sphere of radius $8h^{-1}$Mpc, $T_Q$ --       
the amplitude of the quadrupole component of the CMB anisotropy,       
and the correlation function of observed or simulated       
matter distribution. Some other methods were discussed in DDMT. 

The amplitude of initial perturbations, $A$, is simply linked 
with $\sigma_8$
\be
\sigma_8^2 = 9A^{2}\int_0^\infty dx x^3T^2(x)\left({\sin x_8-x_8
\cos x_8\over x_8^3}\right)^2\,,
\label{as8}
\ee
where $x=k/k_0$, $x_8=8x\Omega_{m}h^{2}$ and $T(x)$ was introduced 
in Sec. 2 . Using the latest estimates (Spergel et al. 2003) for 
the $\Lambda$CDM model (\ref{B1}) we get: 
\be
\sigma_8\approx 0.055A \approx 0.9\pm 0.1,\quad A\approx 16.4\pm 
1.82\,, 
\label{s8}
\ee
\[
\tau_0=\tau_8\approx (0.22\pm 0.02)\,.
\]  
     
Using results of Bunn~\&~White (1997), we can link the amplitude        
$\tau_0$ with the quadrupole anisotropy of the CMB, $T_Q$.        
For the same basic model we have       
\be       
\tau_0=\tau_T\approx 0.27~\left(h\over 0.65\right)^{0.8}       
\left({\Omega_{m}h^{2}}\over 0.2\right)^{1.2}{T_Q\over 20\mu K}\,,       
\label{bw}       
\ee       
Similar expression can be also written for hyperbolic cosmological        
models (DD99).       
       
The variance of displacement (7), $\sigma_s$ and $\tau$, can be       
directly expressed through the observed two point correlation       
function of galaxies, $\xi_{gal}(r)$, as follows:       
\[       
\sigma_s^2 = \lim_{r\rightarrow\infty}       
\int_0^r dx \left(1-{x\over r}\right)x\xi_{gal}(x)\,,       
\]       
and for the correlation function $\xi(r)$ approximated       
by the power law,        
\[       
\xi_{gal}(r) = (r_0/r)^\gamma,\quad r\leq r_\xi\,,       
\]       
we have       
\be       
\sigma_s^2\approx {r_\xi^{2-\gamma}r_0^\gamma\over        
(2-\gamma)(3-\gamma)}\,.       
\label{sss}       
\ee       
Here $r_\xi$ is the first zero-point of the correlation       
function. It is usually found with a small precision, but for       
$\gamma\approx$ 1.5 -- 1.7, $1-\gamma/2\approx$ 0.25 -- 0.15,       
even essential variations of $r_\xi$ do not change significantly       
the final estimates of $\sigma_s$ and $\tau$. Evidently, the 
nonlinear clustering of galaxies at small $r$ increases the 
estimate (\ref{sss}) of $\sigma_s$. However, analysis of 
simulations (DFTT) shows that this expression underestimates 
the amplitude of perturbations.       
       
The correlation functions for the APM survey were found in Loveday    
et al. (1995), for the Las Campanas Redshift Survey in Jing et al.    
(1998) and for the Durham/UKST Redshift Survey in Ratcliffe et al.    
(1998). Using these results we have for $r_\xi\approx (40\pm 10)       
h^{-1}$Mpc that      
\be       
\tau_0=\tau_\xi\approx (0.2\pm 0.04)
\left({{\Omega_{m}h^{2}}\over 0.2}\right)\,.       
\label{adl}       
\ee       
       
Applying the  relations (\ref{ww}) to the systems of walls        
selected in the LCRS, Durham/UKST Redshift Survey, and the SDSS EDR       
(Doroshkevich et al. 2002) allows one to estimate  $\tau_0$ as        
\be       
\tau_0=\tau_w= (0.27\pm 0.04)\sqrt{\Omega_{m}h^{2}/0.2}\,,       
\label{tauw}       
\ee       
what is consistent with the above estimate.       
        
Differences between estimates of $\tau_T$, $\tau_\xi$ and $\tau_w$        
given by (\ref{bw}, \ref{adl}, \ref{tauw}) reflect the precision        
actually achieved in modern observations and, in particular, 
indicate a possible influence of cosmic variance, nonlinearity and 
random factors.        
       
\subsection{Mass function of structure elements} 

The Press--Schechter formalism derived for the spherical 
collapse describes quite well the mass functions of various 
observed and simulated structure elements. In spite of this, 
all attempts to extend it to the collapse of 
asymmetric objects failed with the exception of 
recently proposed  description of collapse of
elliptical clouds (see, e.g., Loeb 
and Barkana 2001; Sheth and Tormen 2002).
In this paper we demonstrate that the Zel'dovich 
theory successfully describes collapse of any object.

Indeed, formation and relaxation of pancakes along the axis 
of the most rapid compression does not prevent their deformation 
in transverse directions due to relatively small gradients of 
density and pressure in these directions. The growth of density 
within the pancake changes the rate of evolution and accelerates 
its compression and further transformation into high density 
filaments and clouds. However, the masses of filaments and 
clouds formed due to such compression remain almost unchanged. 
The same factor also decelerates the expansion of the pancake and 
its dissipation and, so, increases the fraction of surviving 
pancakes. 

This means that, allowing for these deformations, we obtain 
approximate time dependent mass functions of structure 
elements formed due to successive compression in one, two 
and three directions. These results emphasize the generic 
character of the formation of all structure elements and 
link the fundamental characteristics of structure with the 
initial power spectrum. They extend the Press--Schechter 
formalism for all LSS elements including the filaments 
and walls which are far from equilibrium. The unexpected 
similarity of the mass functions for these elements verifies 
that the shape of collapsed clouds influences the rate of 
collapse but it does not change significantly their mass 
functions. 

As is seen from the comparison of results presented in Sec. 
5 and in Loeb \& Barkana (2001) for the Press--Schechter 
formalism, the mass functions derived in both approaches are 
similar and, in particular, the relations (\ref{mnm1}) and 
(\ref{ps}) resemble each other. Both mass functions are 
sensitive to the damping of small scale perturbations 
caused by the random motion of DM particles and the 
Jeans damping. They predict the existence of numerous low 
mass objects which can be identified with isolated dwarf 
galaxies and a rich population of Ly-$\alpha$ absorbers. 
Both approaches predict a strong suppression of formation 
of isolated low mass objects with $M\leq M_\rho$. 

This comparison indicates that the differences between these 
approaches are quantitative rather than qualitative. The 
similarity of the mass functions demonstrates, in fact, the 
self similar character of the process of structure evolution 
that is the successive condensation of matter within clouds, 
filaments and walls with progressively increased sizes and 
masses. At the same time, the approximate character of both 
the Zel'dovich theory and the Press--Schechter formalism 
implies that the proposed mass functions are only approximate. 
Detailed analysis of high resolution simulations with application 
of the approach proposed in this paper will 
allow to improve results obtained in Sec. 5. 

Due to their high overdensity above the mean density, both 
filaments and walls are easily detected in the deep galaxy 
surveys using the Minkowski Functional approach (Schmalzing 
et al. 1999; Kerscher 2000) and the well known 
friend--of--friend method generalized in the Minimal 
Spanning Tree technique (Barrow, Bhavsar \& Sonoda 1985; 
van de Weygaert 1991; Doroshkevich et al. 2001, 2002). 
The mass functions of the observed structure elements can 
be compared with the expected ones. For simulations the same 
comparison can be performed at all redshifts what allows to 
trace the expected redshift dependence of these functions. 

In Sec. 5.7 for the first time we compare the expected mass 
functions with simulated ones. In Doroshkevich et al. (2002) 
the expected mass functions are compared with the observed ones 
for the LSS elements selected with various threshold overdensity 
from the SDSS EDR. These results show that both relations 
(\ref{mnm03}) and (\ref{ps}) describe quite well even the mass 
distribution of filaments and walls with a moderate richness. 
Stronger disagreement appears for the richest walls and filaments 
formed due to the process of percolation which is not described 
by the Zel'dovich theory.         

However, the potential of both approaches is limited as they 
cannot describe the final disruption of collapsed clouds what 
leads to formation of numerous low mass satellites of the central 
object and incorporation of filaments and walls into a joint 
network (percolation process). Neither the Press--Schechter 
formalism nor the Zel'dovich theory can describe the impact of 
environment discussed in Sec. 4 and, in particular, the faster 
evolution of clouds accumulated by richer walls as compared with 
isolated clouds. As was discussed in Sec. 5, both approaches do 
not describe the coagulation of high density clouds what distorts 
their mass function and makes it similar to the power low.  
These problems remain open for further investigations. 
   
\subsection{Interaction of large and small scale perturbations.        
Large scale bias}       
       
The analysis of redshift distribution of absorbers shows        
almost homogeneous spatial distribution of baryonic and DM 
components of the matter on scales $\geq 1 h^{-1}$Mpc. At 
the same time, the observed spatial distribution of luminous 
matter is strongly nonhomogeneous. Thus, about half of the 
observed galaxies are concentrated within large walls while 
within the B\"ootes void the number of galaxies is very small. 
The observed walls and voids are associated with compressed 
and expanded regions of the Universe and these observations 
point out to the possible correlations between the galaxy 
formation and the large scale deformation field (see, e.g., 
Rees 1985; Dekel \& Silk 1986; Dekel \& Rees 1987). Some 
of such correlations were already noticed in simulations 
(see, e.g., Sahni et al. 1994).        
       
As was shown in Sec. 4, the Zel'dovich theory allows to 
quantify this interaction and demonstrates that the formation 
rate of high density objects is modulated by large scale 
perturbations. This modulation cannot significantly change 
the redshift evolution of the fraction of mass accumulated by 
structure elements. However, this interaction accelerates the 
formation of high density halos and galaxies within regions 
associated with the future LSS elements such as clusters of 
galaxies, filaments and walls. These results suggest that the 
poorer sample of isolated galaxies and invisible DM halos 
situated within low density regions were formed later then 
similar galaxies and DM halos situated within filaments and 
walls. The results presented in Sec. 4.1 illustrate this 
influence for simulated matter distribution.

Another important factor which can not be neglected is the 
acceleration of merging and 
coagulation processes within filaments and walls. Both 
processes change the mass function of observed galaxies 
and increase the fraction of massive objects. The high 
concentration of observed galaxies within filaments and 
walls and small fraction of  isolated galaxies    
($<~10\%$) is an evidence in favor of the large scale 
bias. These factors also determine the observed correlation 
function of galaxies. 

\subsection{Properties of low mass pancakes and the mass 
of DM particles}       
       
Numerous simulations performed recently (see, e.g., Zhang et al.    
1998;, Weinberg et al. 1999) demonstrate that the absorption    
lines observed in spectra of the farthest quasars -- the 
Ly--$\alpha$ forest -- are related to the numerous low mass 
clouds formed at high redshifts. As was discussed in 
Demia\'nski et al. (2003), some of the statistical 
characteristics of observed absorbers can be successfully 
explained on the basis of Zel'dovich theory. At the same time, 
the analysis shows that the approximate description based on 
results obtained in DD99 have to be essentially improved. 
Indeed, the absorbers are observed in a wide range of redshifts 
and the properties of long--lived absorbers are changing with 
time at least due to their transverse expansion and compression. 

Discussion of this problem in Sec. 7 shows that for 
statistically homogeneous sample of absorbers we can neglect 
this influence, at least for their two important characteristics, 
namely, for the PDF of their surface density
(\ref{ww}) and their mean linear number density
(\ref{nq1}). Results obtained in Sec. 7 link these 
characteristics with the properties of 
the initial power spectrum at small scale.
This means that the analysis of these 
characteristics allows one to measure the variance of initial 
density perturbations and to restrict the mass of dominant 
fraction of dark matter particles.        
       
Results obtained in Sec. 7 we applied to
the Lyman-$\alpha$ forest 
in Demia\'nski, Doroshkevich \& Turchaninov (2003). This analysis 
confirms that the relations (\ref{ww}) and (\ref{nq1}) describe quite 
well the PDF and the redshift distribution of $\sim$ 5000 observed 
absorbers. These results also verify the Gaussianity of initial 
perturbations. In turn, analysis of redshift distribution of 
absorbers allows one to estimate the actual characteristics of 
the initial power spectrum as follows:
\be
q_0\approx 0.01\pm 0.003,~ m_0\approx 0.2-0.5,~ 
M_{DM}\geq 1-5{\rm keV}.
\ee
Analysis of simulations (Narayanan et al. 2000; Barkana, 
Haiman \& Ostriker 2001) restricts the mass of DM particles 
to  $M_{DM}\geq$ 1--1.5 keV. 
Comparison of other characteristics of pancakes derived in 
Secs. 3\,\&\,7 with observations can be found in Demia\'nski, 
Doroshkevich \& Turchaninov (2003). 

\subsection{Reheating of the universe}       
       
The relations (\ref{wcl}, \ref{mnmr}) show that in the Zel'dovich       
theory of gravitational instability the rate of matter collapse 
at high redshifts depends upon both the amplitude of perturbations,        
$\tau(z)$, as given by (\ref{tau}), and the coherent length of        
density field, $l_\rho~\&~q_0$, as given by (\ref{lv}), which, in        
turn, depends on the shape of the initial power spectrum at large        
$k$ and the mass of the dominant type of DM particles.        
       
As was noticed in Sec. 3, the most effective matter condensation        
within high density clouds occurs at $\tau(z_r)=\tau_r       
\approx\sqrt{q_0/6}$ (see equation (\ref{zr})) when these clouds 
already  accumulate $\sim$ 2\% of matter and the main fraction        
of mass is concentrated within clouds with        
$M\sim (0.2 - 0.5)\langle M\rangle$.        
       
For $q_0\sim 10^{-2}$ we can estimate the redshift of the period 
of most efficient condensation and the mean mass of DM clouds as 
follows:       
\[       
\tau_r\sim 0.04,\quad 1+z_r\approx 1.3\tau_0/\tau_r\sim 7 - 8\,,        
\]       
\be       
\langle M_{DM}\rangle\sim \tau_r^3 M_v\approx 10^{12}M_\odot       
\left({0.3\over\Omega_m}\right)^2\left({0.7\over h}\right)^4\,,       
\label{q100}       
\ee       
what is similar to the mass of a typical galaxy. For $q_0\sim        
10^{-3}$ we have, respectively,       
\[       
\tau_r\sim0.013,\quad 1+z_r\approx 1.3\tau_0/\tau_r\sim 20\,,       
\]        
\be       
\langle M_{DM}\rangle\sim4\cdot 10^{-2}\tau_r^{3}M_v
\approx 10^9M_\odot       
\left({0.3\over\Omega_m}\right)^2\left({0.7\over h}\right)^4\,,       
\label{q1000}       
\ee       
what is similar to the mass of a typical dwarf galaxy. These 
results illustrate strong links between the period of reheating 
and the shape of the power spectrum  at small scale characterized 
by $q_0$. Even at higher redshifts noticeable fraction of matter 
can be compressed within high density clouds with masses of 
galactic scales.       
       
According to available estimates (Haiman \& Loeb 1999; Loeb and        
Barkana 2001) the reionization of the universe occurs after     
$\sim$ 1 - 3\% of matter is concentrated within high density halos. 
From (\ref{q100}, \ref{q1000}) it follows that this fraction of 
collapsed matter can be reached already at $\tau\sim\tau_r$ at 
redshifts $z= z_r\sim$ 8 -- 25, for $q_0\sim 10^{-2} - 10^{-3}$. 
The effective reionization and reheating of the universe at such 
redshifts is consistent with  the observed concentration of neutral 
hydrogen at redshifts $z\sim$ 6, which characterizes mainly the rate 
of generation and achieved intensity of UV background. This means 
that these ranges of $q_0$ and $z_r$ provide a reasonable 
estimate of the period of reheating.        
        
The observations of environment of high redshift quasars (Fan et        
al. 2001) provide an evidence in favor of reheating of the Universe        
at $z\sim$ 6        
-- 10, what is more consistent with $q_0\sim 10^{-2}$.        
However, our analysis shows that, for the standard WDM model with        
the Harrison -- Zel'dovich large scale power spectrum and exponential        
cutoff caused by a mass $M_{DM}\sim$ 1 keV of the dominant fraction      
of DM particles, some problems can appear with formation of         
low mass isolated galaxies with $M\leq 10^8 - 10^9 M_{\odot}$.
Perhaps, some excess of power at small scales can help to      
solve this problem. First observational indications of such excess 
are presented in Demia\'nski\,\&\,Doroshkevich (2003) 
       
\subsection*{Acknowledgments}       
AD thanks V.Lukash and E.Mikheeva for useful discussions.        
This paper was supported in part by Denmark's Grundforskningsfond       
through its support for an establishment of Theoretical Astrophysics       
Center and a grant of the Polish State Committee for Scientific 
Research AD also wishes to acknowledge support from the Center       
for Cosmo-Particle Physics "Cosmion" in the framework of the project       
"Cosmoparticle Physics".

\appendix
\section{Correlation functions of initial perturbations.}       
       
In this appendix we present a few correlation and structure       
functions which describe the relative spatial distribution       
of important parameters of the initial perturbations. These functions       
have been introduced in DD99, where more details can be found.       
       
The structure function of gravitational potential perturbations       
characterizes correlation of the gravitational potential in two       
points $\tilde{\bf q}_{1}$ and $\tilde{\bf q}_{2}$. As the power       
spectrum is  a function of only the absolute value of  wave number       
$|k|$, this structure function depends on $\tilde{q}_{12}=       
|\tilde{\bf q}_{1}- \tilde{\bf q}_{2}|$ and for the perturbations       
of gravitational potential we have       
\be       
3{<\Delta\phi_{12}\Delta\phi_{34}>\over l_v^2\sigma_s^2} =       
G_0(\tilde{q}_{14})-G_0(\tilde{q}_{13})+
\label{A1}       
\ee
\[G_0(\tilde{q}_{23})-G_0(\tilde{q}_{24}),\quad
\tilde{q}_{ij}=|\tilde{\bf q}_i-\tilde{\bf q}_j|\,,       
\]       
\[       
\Delta\phi_{12} = \phi(\tilde{\bf q}_1)-\phi(\tilde{\bf q}_2),       
\quad \Delta\phi_{34} = \phi(\tilde{\bf q}_3)-\phi
(\tilde{\bf q}_4)\,,       
\]       
\[       
G_0(\tilde{q}) = {3\over \pi^2l_v^2\sigma_s^2}\int_0^\infty {p(k)       
\over k^2}\left(1-{\sin{k\tilde{q}}\over k\tilde{q}}\right)dk\,.       
\]       
Differentiations of this structure function give other structure       
functions:       
\[       
{3<\Delta\phi(\tilde{\bf q}_{12}) S_i(\tilde{\bf q}_3)>\over l_v       
\sigma_{s}^2}= 
\]\[(\tilde{\bf q}_1-\tilde{\bf q}_3)_iG_1(\tilde{q}_{13})       
-(\tilde{\bf q}_2-\tilde{\bf q}_3)_iG_1(\tilde{q}_{23})\,,       
\]       
\[       
{3<S_i(\tilde{\bf q}_1)S_j(\tilde{\bf q}_2)>\over\sigma_{s}^2}=       
\delta_{ij}G_1(\tilde{q}_{12})+(\tilde{\bf q}_{12})_i
(\tilde{\bf q}_{12})_jG_{2}(\tilde{q}_{12})\,,       
\]       
\be       
G_1(x)= G_0'/x,~~G_2(x)= G_1'/x,~~G_3(x)= G_2'/x\,,       
\label{A2}       
\ee       
\[       
G_{12}(x)=G_1(x)+x^2 G_2(x)=G_0^{''} = (xG_1)^{'}\,,       
\]       
       
For the Harrison - Zel'dovich primordial power spectrum and       
for the CDM transfer function (BBKS), these functions can be       
approximated by       
\be       
G_1(x)\approx (1+q_0)[1+\sqrt{q_0^2+x^2}+a_0x^2]^{-1}\,,       
\label{A.3}       
\ee       
\[       
G_{12}\approx G_1^2(1+q_0)^{-1}[1-a_0x^2+q_0^2/\sqrt{q_0^2+x^2}]\,,       
\]       
\[     
G_1(0) = G_{12}(0) = 1,\quad G_{2}(0) = -[q_0(1+q_0)]^{-1}\,.     
\]     
\[       
x=\tilde{q}/l_v,~~q_0 = l_\rho/l_v,~~a_0 = 5~(l_v/L_0)^2\approx 0.3\,,       
\]       
where the typical scale $L_0$ is defined as       
\[       
L_0^2 = 3 \pi^2\int_0^\infty p(k)\left(1-{\sin{kL_0}       
\over kL_0}\right)k^{-2}dk\Big/\int_0^\infty p(k) dk\,,       
\]       
and the scales $l_v \& q_0$ were introduced by (\ref{lv}).     

\section{Characteristics of the deformation field.}       
       
The general deformation of a spherical cloud with a diameter       
$q=\tilde{q}/l_v$ can be suitably characterized by the       
dimensionless random scalar function       
\be       
\Theta({\bf q})=\Theta(q,\theta,\phi) =       
{\sqrt{3}[{\bf S}(\tilde{\bf q}/2)-{\bf S}(\tilde{\bf -q}/2)]       
\cdot {\bf q}\over\sigma_s q}\,,       
\label{B.1}       
\ee       
instead of the deformation tensor $d_{ik}$. Expansion of this       
function into spherical harmonics characterizes deformations       
of the cloud with a required accuracy. This problem is similar       
to the usual description of CMB anisotropy in terms of spherical       
harmonics, and we have:       
\be       
\Theta(q,\theta,\phi) = \sum_{l,m}~a_{lm}(q)Y_{l,m}(\theta,\phi)\,,       
\label{B.2}       
\ee       
\[       
Y_{l,m}(\theta,\phi) = (-1)^mi^l\sqrt{ {2l+1\over 4\pi}       
{(l-m)!\over(l+m)!} }P_l^m(\cos\theta)e^{im\phi}\,,       
\]       
\[       
Y_{l,-|m|}(\theta,\phi) = (-1)^{l-m}Y_{l,m}(\theta,\phi),       
~l\geq m\geq 0\,,       
\]       
\[       
\langle\Theta({\bf q}_1)\Theta({\bf q}_2)\rangle =       
\Psi(q,\mu_{12}) =\sigma^2(q)\sum_l b_l^2(q)P_l(\mu_{12})=        
\]       
\[       
G_1(x_1)+G_1(x_2)-G_{12}(x_1)-G_{12}(x_2)+\]\[
\mu_{12}[G_1(x_2)-G_1(x_1)+G_{12}(x_1)-G_{12}(x_2)]
\]       
\[       
\mu_{12} ={{\bf q}_1{\bf q}_2\over q_1q_2},~~x_1=q\sqrt{1+       
\mu_{12}\over 2},~ x_2=q\sqrt{1-\mu_{12}\over 2}\,,       
\]       
\be       
\sigma^2(q)b^2_l(q) =\sum_{m=-l}^l{\langle a_{lm}^2\rangle       
\over 4\pi} ={2l+1\over 4\pi}\langle a_{l}^2\rangle,~\sum_l        
b_l^2=1\,,					       
\label{B.3}       
\ee       
\be       
\Psi(q,\mu_{12}=1) = \sigma^2(q),~~\Psi(q,\mu_{12}=0) =       
\sigma^2(q)r_s(q),					       
\label{B.4}       
\ee       
\be       
\sigma^2(q) = 2[1-G_{12}(q)],~~r_s(q)=       
-{q^2\over\sigma^2}G_2\left({q\over\sqrt{2}}\right)\,.       
\label{B.5}       
\ee       
Here $\sigma^2(q)$ and $r_s(q)$ are the dispersion and the       
coefficient of correlation of dimensionless orthogonal displacements,       
and the functions $G_1$, $G_2$ and $G_{12}$ were introduced in Appendix A.       
The correlation coefficient $r_s$ is a weak function of $q$ and for       
$q\ll 1, ~q_0\ll 1$, we have       
\be       
1/3\leq r_s\leq 2^{-3/2}\,,       
\label{B.6}       
\ee       
what is consistent with the usual correlations of the components       
of deformation tensor.        
       
The function $\Psi(q,\mu_{12})$ is symmetrical with respect to the       
replacement $\mu_{12}\rightarrow -\mu_{12}$ and, therefore, $b^2_l=0$       
for odd $l$. Using the relations (\ref{A.3}) we have for the relative    
amplitude of even spherical harmonics, for the most interesting cases    
$q_0\ll q\ll$1 and $q\ll q_0\ll$1, respectively:       
\be       
b_0^2\approx 0.533,~~ b_2^2\approx 0.381,~~ b_4^2\approx       
0.037,~~ b_6^2\approx 0.014, ...		       
\label{B.7}       
\ee       
\be       
b_0^2\approx 0.55,\quad b_2^2\approx 0.44\gg b_4^2, b_6^2 ... .		       
\label{B.8}       
\ee       
These results justify the assumption used in DD99 to neglect       
higher order harmonics of perturbations with $l\geq 4$.        
       
\end{document}